\documentclass[preprint]{ptephy_v1}

\preprintnumber{KUNS-2800} 

\newcommand\ii{\mathrm{i}}

\usepackage{color,url}

\usepackage{braket}
\usepackage[T1]{fontenc}
\newcommand{\bvec}[1]{\mbox{\boldmath $#1$}}

\begin{document}

\title{Signature dependent triaxiality 
for shape evolution from superdeformation 
in rapidly rotating $^{40}$Ca and $^{41}$Ca}

\author{Shinkuro Sakai}
\affil{Graduate School of Science and Technology, Niigata University, Niigata 950-2181, Japan }

\author{Kenichi Yoshida}
\affil{Department of Physics, Kyoto University, Kyoto 606-8502, Japan \email{kyoshida@ruby.scphys.kyoto-u.ac.jp}}

\author{Masayuki Matsuo}
\affil{Department of Physics, Niigata University, Niigata 950-2181, Japan \email{matsuo@phys.sc.niigata-u.ac.jp}}



\begin{abstract}%
We investigate the possible occurrence of the highly-elongated shapes 
near the yrast line in $^{40}$Ca and $^{41}$Ca at high spins on the basis of the 
nuclear energy-density functional method. 
Not only the superdeformed (SD) yrast configuration 
but the yrare configurations on top of the SD band 
are described by solving the 
cranked Skyme-Kohn-Sham equation 
in the three-dimensional coordinate-space representation.   
It is suggested that some of the excited SD bands undergo band crossings and develop to 
the hyperdeformation (HD) beyond $J \simeq 25 \hbar$ in $^{40}$Ca. 
We find that 
the change of triaxiality in response to rotation plays a decisive role for the shape evolution 
towards HD, 
and that this is governed by the signature quantum number of the last occupied 
orbital at low spins. 
This mechanism can be verified in an experimental observation of the positive-parity SD yrast signature-partner bands in $^{41}$Ca, 
one of which ($\alpha=+1/2$) undergoes crossings with the HD band 
while the other ($\alpha=-1/2$) shows the smooth evolution from 
the collective rotation at low spins to the non-collective rotation with oblate shape at the termination.

\end{abstract}

\subjectindex{D11, D13}

\maketitle

\section{Introduction}

Rotational motion, a typical collective mode of excitation in nuclei, emerges to restore the rotational symmetry broken by the nuclear deformation~\cite{BM2}. 
In most cases, a prolate deformation occurs naturally in the ground state, and then 
the total spin is generated by the collective rotation about the axis perpendicular to the symmetry axis. 
On the other hand, the spin is constructed only by the alignment of the single-particle orbitals when the rotation axis coincides with 
the symmetry axis. 
Actually, the system may deviate from the axial symmetry due to the Coriolis effect as soon as it rotates, 
and the direction of 
the angular momentum vector generated by the single-particle orbitals can be different from either the symmetry axis or the rotation axis. 
Therefore, the interplay and coupling between the collective and single-particle motions have to be considered selfconsistently 
to investigate the rotational motions of nuclei~\cite{voi83}.

Superdeformed (SD) states exhibiting beautiful 
rotational spectra are an ideal situation in which the concept of the nuclear deformation is realized, and thus 
provide an opportune playground for the study of rotational motions. 
The study of SD bands has been an active field 
in nuclear physics and the rotational bands 
have been observed up to high spins in various mass regions~\cite{sin02} 
since its discovery in 1986~\cite{twi86}. 
Recently, the high-spin structures in light $N \simeq Z$ nuclei near the doubly-magic $^{40}$Ca nucleus 
have been studied experimentally~\cite{sve00, sve01, rud02, abr19, ide10,sod12, ide01,chi03, bha16, lac03, mor04, ole00},
and SD bands have been observed in such as $^{36}$Ar~\cite{sve00,sve01}, $^{40}$Ar~\cite{ide10}, 
$^{40}$Ca~\cite{ide01,chi03}, $^{42}$Ca~\cite{had16}, and $^{44}$Ti~\cite{ole00}. 
An interesting feature in this mass region is the coexistence of states with different shapes 
in low energy, which is caused by
the single-particle excitations from the core and 
the coherent shell effects of neutrons and protons. 
Complementary theoretical models have been used in attempt to describe microscopically the 
SD state in $^{40}$Ca 
by employing the interacting shell model~\cite{pov03,cau07}, 
the nuclear energy-density functional (EDF) method~\cite{ina02,ben03,ray16}, 
and the various cluster models~\cite{sak04,eny05,tan07}.
The yrast spectroscopy of these light nuclei brings unique opportunity to investigate the mechanism 
for the occurrence of the SD band and the possibility of the hyperdeformed (HD) band 
because the single-particle density of state around the Fermi levels is low 
so that we can study in detail the deformed shell structures 
responsible for the SD and HD bands.

In quest of HD bands, shape evolution from the SD states 
associated with an increase in spin has been investigated 
because 
the nuclear rotation could stabilize the HD state due to its large moment of inertia. 
In fact HD bands as well as SD bands with various configurations are predicted 
in the  cranked Skyrme-Kohn-Sham (SKS) calculation~\cite{ina02}  and
in the cranked relativistic-mean-field (CRMF) calculation~\cite{ray16}, in which
the observed SD band in $^{40}$Ca is reasonably explained. 
An interesting feature in these predictions is that the obtained 
SD states are triaxially deformed: 
The magnitude of triaxial deformation is almost constant $\gamma \sim -10^\circ$ as a function 
of spin in the CRMF~\cite{ide01,ray16} while the SD states obtained in the cranked SKS calculation has 
a slightly smaller triaxiality $\gamma \sim 6$ -- $9^\circ$ changing with spin~\cite{ina02}. 
The antisymmetrized molecular dynamics (AMD) calculation gives the similar result to the ones in the mean-field calculations 
with $\gamma \sim 15^\circ$, 
and the $K^\pi=2^+$ band is predicted to appear due to the triaxial deformation of the SD band~\cite{tan07}. 
The CRMF calculation predicts also that one of the SD bands exhibits a band termination~\cite{ray16}.
We note however that the preceding studies do not clarify how  the triaxial deformation and the termination 
emerge in the SD bands, and how the SD bands change their shape from SD to HD. 

In the present study, we investigate the near-yrast structures 
in $^{40}$Ca and $^{41}$Ca at high spins 
on the basis of the cranked SKS method with paying attention to not only the evolution of shape elongation 
from the SD to HD states but the change of triaxiality in response to rotation. 
A binding energy, a functional of densities, is minimized with the configuration being constrained so that 
the polarization associated with the particle-hole (ph) excitations and the time-odd components in the mean field 
is taken into account.  
The possible emergence of the HD state is explored, and then the microscopic mechanism is discussed. 
For $^{40}$Ca, 
we find that the excited SD bands undergo crossings at $J \simeq 25\hbar$, in which the hyperintruder shell is occupied 
by both a neutron and a proton, forming negative-parity HD bands. 
For realization of the HD states at high spins, the signature quantum number of 
the last occupied low-$\Omega$ orbital near the Fermi level plays a decisive role in connection with 
the development of triaxiality. 
We then propose that this mechanism is verified in observing the SD yrast signature-partner bands in $^{41}$Ca, 
one of which undergoes crossings with the HD band while the other shows the smooth termination from the collective
rotation at low spins to the non-collective rotation at high spins.

The article is organized as follows. 
Section~\ref{sec_cal} describes the calculation scheme for the study of the near-yrast high-spin states 
in the framework of the nuclear EDF method; 
how to constrain and specify the configurations is explained. 
Section~\ref{results} shows the results of the calculation: 
The near-yrast structures in $^{40}$Ca are discussed in Sec.~\ref{res_40Ca}. 
The high-spin states in $^{41}$Ca are investigated in Sec.~\ref{res_41Ca}. 
The obtained results and the underlying mechanism for the occurrence of the HD states at high spins 
are confirmed by performing the calculation with a different EDF in Sec.~\ref{SkI}. 
Finally, Sec.~\ref{summary} summarizes the results of our work. 

\section{Calculation scheme \label{sec_cal}}

We define the $z$-axis as a quantization axis of the intrinsic spin and 
consider the system rotating uniformly about 
the $z$-axis. 
Then, the cranked SKS equation is given by~\cite{bon87}
\begin{equation}
\delta (E[\rho] - \omega_\mathrm{rot}\braket{ \hat{J}_z})=0,
\label{eq:cranking}
\end{equation}
where $E[\rho]$ is a nuclear EDF, 
$\omega_\mathrm{rot}$ and $\hat{J}_z$ mean the rotational frequency and 
the $z$-component of angular momentum operator, 
and the bracket denotes the expectation value with respect to the Slater determinant 
given by the occupied single-particle KS orbitals for a given $\omega_\mathrm{rot}$. 

To discuss the shape of calculated density distribution, it is convenient to introduce 
the multipole moments given as
\begin{equation}
    \alpha_{lm}=\frac{4\pi}{3A\bar{R}^l}\int d^3r\ r^lX_{lm}(\hat{r})\rho(\bvec{r}),
  \label{eq:MultMome} 
\end{equation}
where $\rho(\bvec{r})$ is the particle density, 
$\bar{R}=\sqrt{\tfrac{5}{3A}\int d^3r\ r^2\rho(\bvec{r})}$, and 
$X_{lm}$ are real basis of the spherical harmonics 
\begin{equation}
  X_{lm}(\hat{r})=
  \begin{cases}
    Y_{l0}(\hat{r}) & (m=0)\\
    \tfrac{1}{\sqrt{2}}[Y_{l,-m}(\hat{r})+Y^\ast_{l,-m}(\hat{r})] & (m>0) \\
    \tfrac{1}{\sqrt{2}i}(-1)^m[Y_{lm}(\hat{r})-Y^\ast_{lm}(\hat{r})] & (m<0).
  \end{cases}
\end{equation}
We then define the quadrupole deformation parameter $\beta$ and the triaxial deformation parameter $\gamma$ by
\begin{equation}
  \alpha_{20}=\beta\cos\gamma,\ \ \ \alpha_{22}=\beta\sin\gamma.
\end{equation}

Since we defined the $z$-axis as a rotation axis, 
the spherical harmonics is redefined 
in the transformation $(x,y,z)\to(z,x,y)$, 
and the expressions needed for describing the quadrupole deformations are given as
\begin{equation}
  \begin{cases}
    Y_{20}=\sqrt{\frac{5}{16\pi}}\left(\frac{3z^2}{r^2}-1\right)\\
    Y_{22}=\sqrt{\frac{15}{16\pi}}\frac{x^2+y^2}{r^2}\\
    \frac{Y_{2,+2}+Y_{2,-2}}{\sqrt{2}}=\sqrt{\frac{15}{16\pi}}\frac{x^2-y^2}{r^2}\\
  \end{cases}\to
  \begin{cases}
    Y_{20}=\sqrt{\frac{5}{16\pi}}\left(\frac{3y^2}{r^2}-1\right)\\
    Y_{22}=\sqrt{\frac{15}{16\pi}}\frac{z^2+x^2}{r^2}\\
    \frac{Y_{2,+2}+Y_{2,-2}}{\sqrt{2}}=\sqrt{\frac{15}{16\pi}}\frac{z^2-x^2}{r^2}.
  \end{cases}\
\end{equation}

In the numerical calculations, we impose the reflection symmetry about the $(x, y)$-, $(y,z)$- and $(z, x)$-planes. 
We can therefore construct the simultaneous engenfunctions of the parity transformation $\hat{P}$ 
and the $-\pi$ rotation about the $z$-axis $\hat{R}_z=e^{\ii \pi \hat{j}_z/\hbar}$:
\begin{align}
    \hat{P}\phi_k&=\pi_k\phi_k, \\
    \hat{R}_z\phi_k&=r_k\phi_k,
\end{align}
besides the cranked-KS equation
\begin{equation}
\hat{h}^\prime \phi_k = \epsilon_k \phi_k, 
\label{eq:cSKS}
\end{equation}
with the single-particle Hamiltonian, or the Routhian for $\omega_\mathrm{rot} \ne 0$, namely 
$\hat{h}^\prime=\frac{\delta E}{\delta \rho} - \omega_\mathrm{rot}\hat{j}_z$. 
The eigenvalues $\pi_k$ $(=\pm 1)$ and $r_k$ ($=\pm \ii$) are called 
the parity and $z$-signature, respectively. Hereafter, we simply call the latter signature. 
We can introduce the signature exponent quantum number $\alpha$ ($=\pm 1/2$) by $r \equiv e^{\ii \pi \alpha}$. 
The signature exponent $\alpha$ is useful when comparing with the experimental data through 
the relation $\alpha=I \mod 2$, where $I$ is the total nuclear spin~\cite{voi83}.

Since we chose the quantization axis of the intrinsic spin to coincide with the rotation axis ($z$-axis), 
we can determine the single-particle wave functions such that they satisfy the following reflection symmetries~\cite{bon87, oga09}: 
\begin{align}
    \phi_{k}(-x,y,z,\sigma)&=-\ii r_k\sigma\phi^\ast_{k}(x,y,z,\sigma),   \label{eq:WFsym1}\\
    \phi_{k}(x,-y,z,\sigma)&=\phi^\ast_{k}(x,y,z,\sigma),   \label{eq:WFsym2}\\
    \phi_{k}(x,y,-z,\sigma)&=-\ii \pi_kr_k\sigma\phi_{k}(x,y,z,\sigma)   \label{eq:WFsym3},
\end{align}
with $\sigma$ being the direction of the intrinsic spin. 
We solve Eq.~(\ref{eq:cSKS}) 
by diagonalizing the single-particle Routhian $\hat{h}^\prime$
in the three-dimensional Cartesian-mesh representation with the box boundary condition. 
Thanks to the reflection symmetries (\ref{eq:WFsym1})-(\ref{eq:WFsym3}), 
we have only to consider explicitly the octant region in space 
with $x\ge0$, $y\ge0$, and $z\ge0$.
We use a 3D lattice mesh $x_i=ih-h/2, y_j=jh-h/2, z_k=kh-h/2 \ \  (i,j,k=1,2,\cdots)$ with a mesh size $h=0.8$ fm and 12 points for each direction. 
The differential operators are represented by use of the 9-point formula of finite difference method.
For diagonalization of the Routhian, we use the LAPACK {\sc dsyevx} subroutine~\cite{LAPACK}. 
A modified Broyden's method~\cite{bar08} is utilized to calculate new densities during the selfconsistent iteration.

In the present calculation, we employ the Skyrme EDFs for $E[\rho]$ in Eq.~(\ref{eq:cranking}).
All the time-even densities are included, while the coupling constants for the derivative of spin density 
are set to zero as in Ref.~\cite{bon87} to avoid the numerical instability~\cite{les06,hel12}. 
In some EDFs such as the SkI series~\cite{rei95}, the center-of-mass correction is considered 
by subtracting 
$\langle \hat{\boldsymbol{P}}^2_\mathrm{CM}\rangle/2mA$. 
We, however, take into account the center-of-mass correction simply by $\sum_i\langle \hat{\boldsymbol{p}}^2_i \rangle/2mA$ 
because we do not discuss the total binding energy but rather the relative energy.

Single-particle orbitals are labeled by $[N n_3 \Lambda]\Omega (r)$ with 
$[N n_3 \Lambda]\Omega$ being the asymptotic quantum numbers (Nilsson quantum numbers) 
of the dominant component of the wave function at 
$\omega_\mathrm{rot}=0$ and $r$ the signature of the orbital. 
To describe various types of rotational bands under the energy variation, 
the Slater determinantal states are constructed by imposing the configuration of the single-particle KS orbitals. 
Since the parity and signature are a good quantum number, 
and the pairing correlations are not included in the present calculation, 
the intrinsic configurations of interest can be
described by the occupation number of particle $n$ for the orbitals specified by the quantum number $(\pi, r)$; 
$[n_{(+1, +\ii)} n_{(+1,-\ii)}n_{(-1, +\ii)}n_{(-1, -\ii)}]_q$ for $q=\nu$ (neutron) and $\pi$ (proton) 
as in the cranking calculation code {\sc hfodd}~\cite{HFODD}.

The procedure of the cranking calculation is as follows: 
For a specified intrinsic configuration,
we first find a selfconsisitent deformed solution at zero rotational frequency 
$\omega_\mathrm{rot}=0$ (or at a finite value of $\omega_\mathrm{rot}$ ) 
with use of an initial trial state, which is given by 
a deformed Woods-Saxon potential or  another deformed SKS solution.
We then increase slightly the rotational frequency by 
$\Delta\omega_\mathrm{rot}=0.1$ MeV$/\hbar$, and obtain the selfconsistent
solution. Repeating this with a gradual increase of the rotational frequency, 
we trace evolution of the rotating deformed state as a function of $\omega_\mathrm{rot}$.
Note that no constraint on shape is imposed, except the reflection symmetry, 
in performing the cranking calculation.

\section{Results and discussion \label{results}}

\subsection{Superdeformation and hyperdeformation in $^{40}$Ca \label{res_40Ca}}

\begin{figure}[t]
\begin{center}
\includegraphics[scale=0.56]{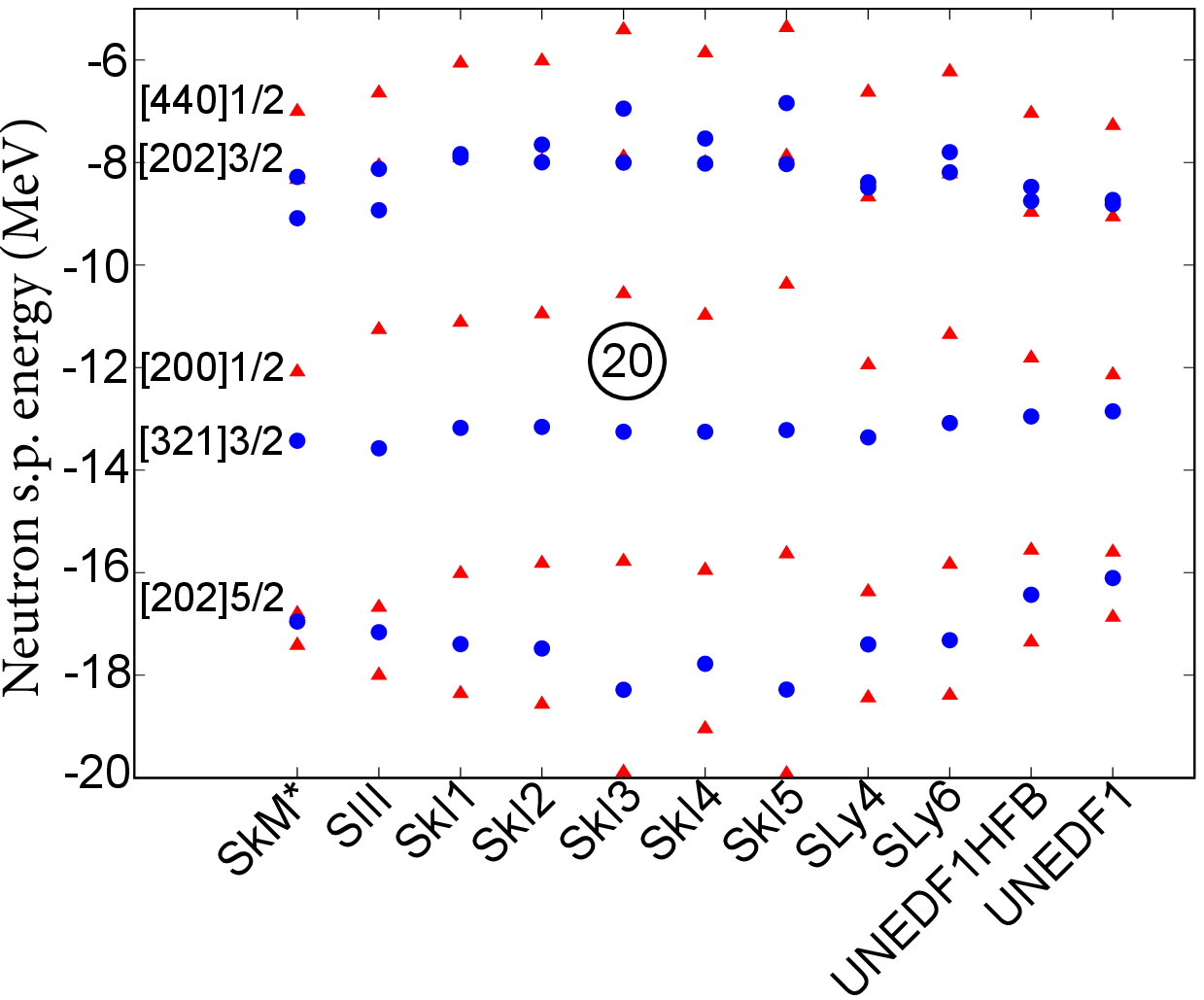}
\includegraphics[scale=0.56]{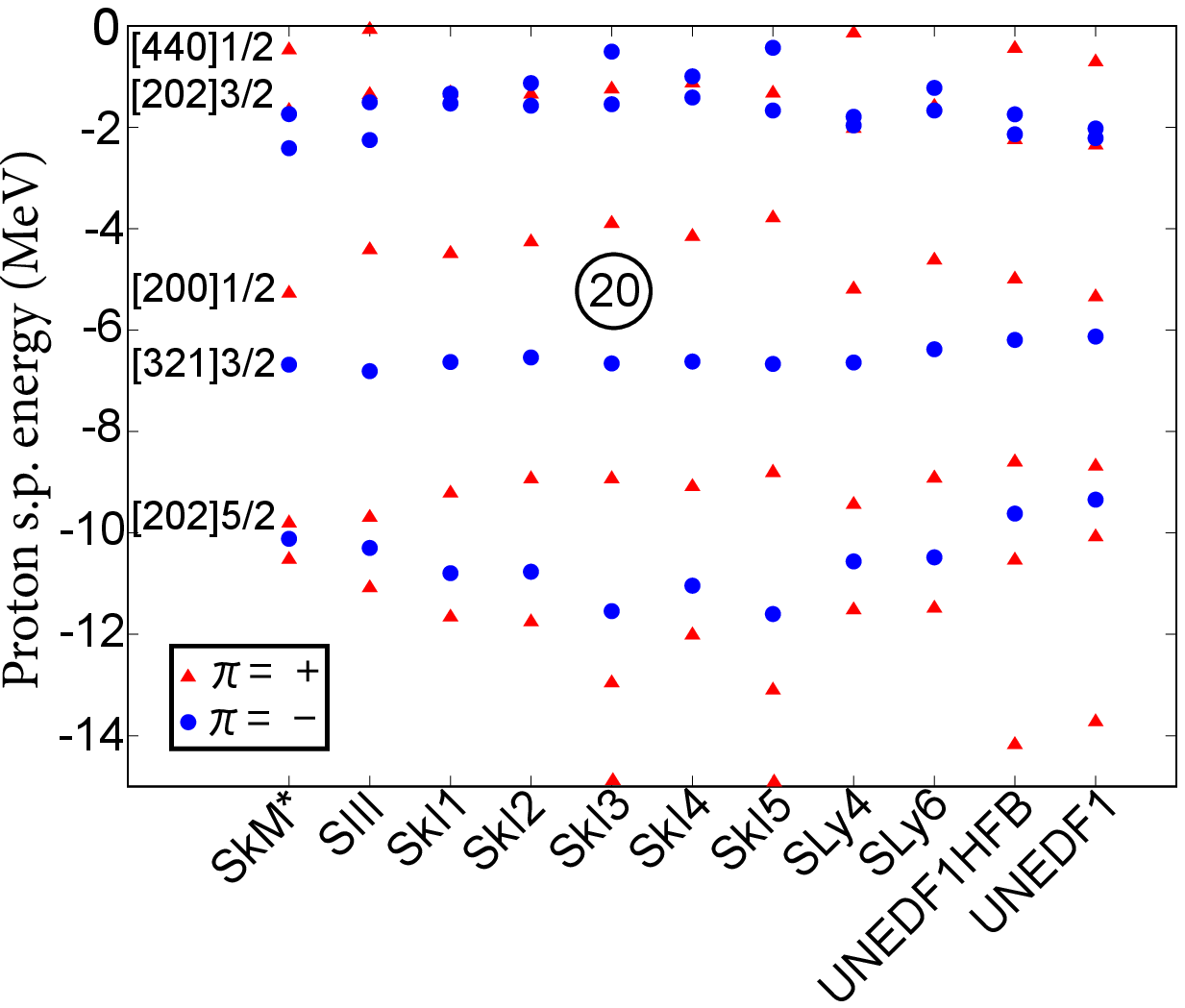}
\caption{
Calculated single-particle energies of neutrons (left) and protons (right) 
for the reference SD configuration $[5555]_\nu[5555]_\pi$ in $^{40}$Ca 
with the use of the several Skyrme EDFs. Triangles and circles represent 
the positive and negative-parity states, respectively. Some orbitals relevant to 
the discussion are labeled by the Nilsson quantum number. 
}
\label{40Ca_SD_spe}
\end{center}
\end{figure}

The ground state of the $^{40}$Ca nucleus, composed of twenty neutrons and twenty protons, 
is calculated to be spherical within the present calculation due to the spherical magic number of $20$ 
generated by a gap between the $d_{3/2}$ and $f_{7/2}$ shells. 
In the present calculation scheme, the spherical configuration is represented as $[7733]_\nu [7733]_\pi$. 
With an increase in the prolate deformation, the [202]3/2 orbital stemming from the spherical $d_{3/2}$ shell 
grows up in energy, and intersects with the down-sloping [330]1/2 orbital originating from the spherical $f_{7/2}$ shell 
(see Fig.~1 in Ref.~\cite{yos05} for example). 
After the crossing of these orbitals, one sees an energy gap that corresponds 
to a normal-deformed (ND) configuration. 
The ND configuration is thus represented as $[6644]_\nu [6644]_\pi$.
When the system is further deformed, an extruder [200]1/2 and an intruder [321]3/2 orbitals cross, 
and an energy gap appears after the crossing of these orbitals, 
leading to an SD configuration 
represented as $[5555]_\nu [5555]_\pi$.
This SD configuration is taken as a reference in the following discussion.  
The ND and SD configurations may correspond to the four-particle four-hole (4p4h) and 
eight-particle eight-hole (8p8h) configurations with respect to the 
spherical configuration, respectively. 
However, the Nilsson-type deformed wave functions at large deformation 
are not simply represented by the orbitals within the single major shell, but 
are expressed by the linear combination of 
the spherical orbitals with admixture of different major shells of $\Delta N=2$;
the 4p4h or 8p8h configuration in terms the spherical shell model may not directly correspond to 
the ND or SD configuration in the present model. 
Furthermore, 
the low-lying $0^+$ states located at 3.35 MeV and 5.21 MeV, 
which are interpreted as the ND and SD states, respectively, in Ref.~\cite{ide01}, 
are actually the mixture of many-particle many-hole configurations 
as described microscopically by the interacting shell model~\cite{cau07}, the generator coordinate method~\cite{ben03}, 
and the AMD~\cite{eny05}.
Therefore, we use somewhat loosely the $n$-particle $n$-hole configuration below. 
Note that even the 12p12h configuration, 
composed of the particles occupying the $g_{9/2}$ shell instead of the $d_{3/2}$ shell
corresponding to the megadeformed (MD) configuration, is also labeled by $[5555]_\nu [5555]_\pi$.

We show in Fig.~\ref{40Ca_SD_spe} the calculated single-particle energies for the reference SD configuration $[5555]_\nu [5555]_\pi$ 
obtained by employing the several Skyrme functionals, 
including SkM*~\cite{bar82}, SIII~\cite{bei75}, SLy4~\cite{cha98}, SLy6~\cite{cha98}, 
SkI1~\cite{rei95}, SkI2~\cite{rei95}, SkI3~\cite{rei95}, SkI4~\cite{rei95}, SkI5~\cite{rei95}, 
UNEDF1-HFB~\cite{sch15}, and UNEDF1~\cite{kor12}.
One sees that the SkI series produce the pronounced SD gap energy at a particle number 20. 
Note that the CRMF calculation in Ref.~\cite{ray16} also gives a high SD-gap energy of $\sim 4$ MeV. 

\begin{figure}[t]
\begin{center}
\includegraphics[scale=0.36]{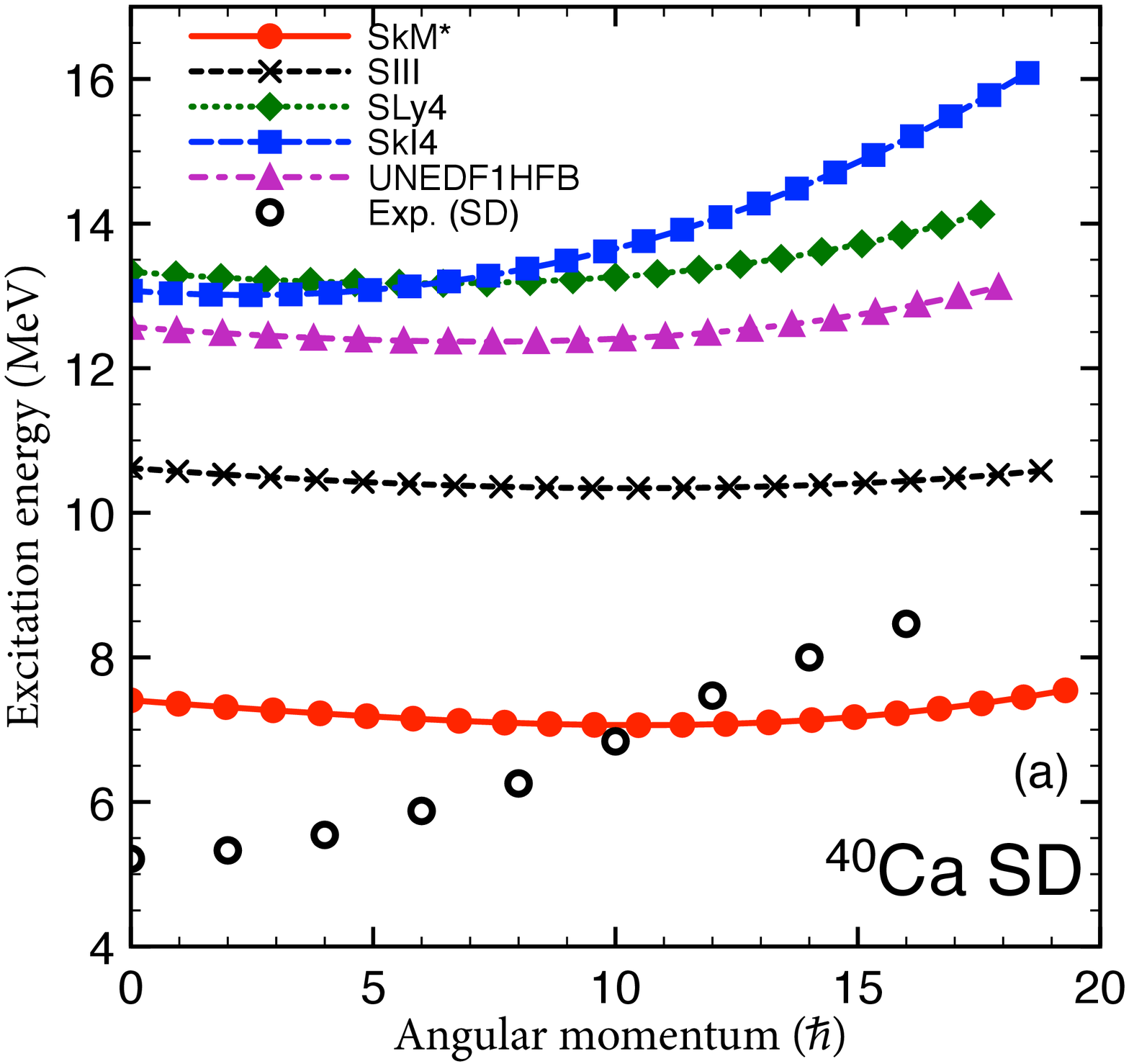}
\includegraphics[scale=0.36]{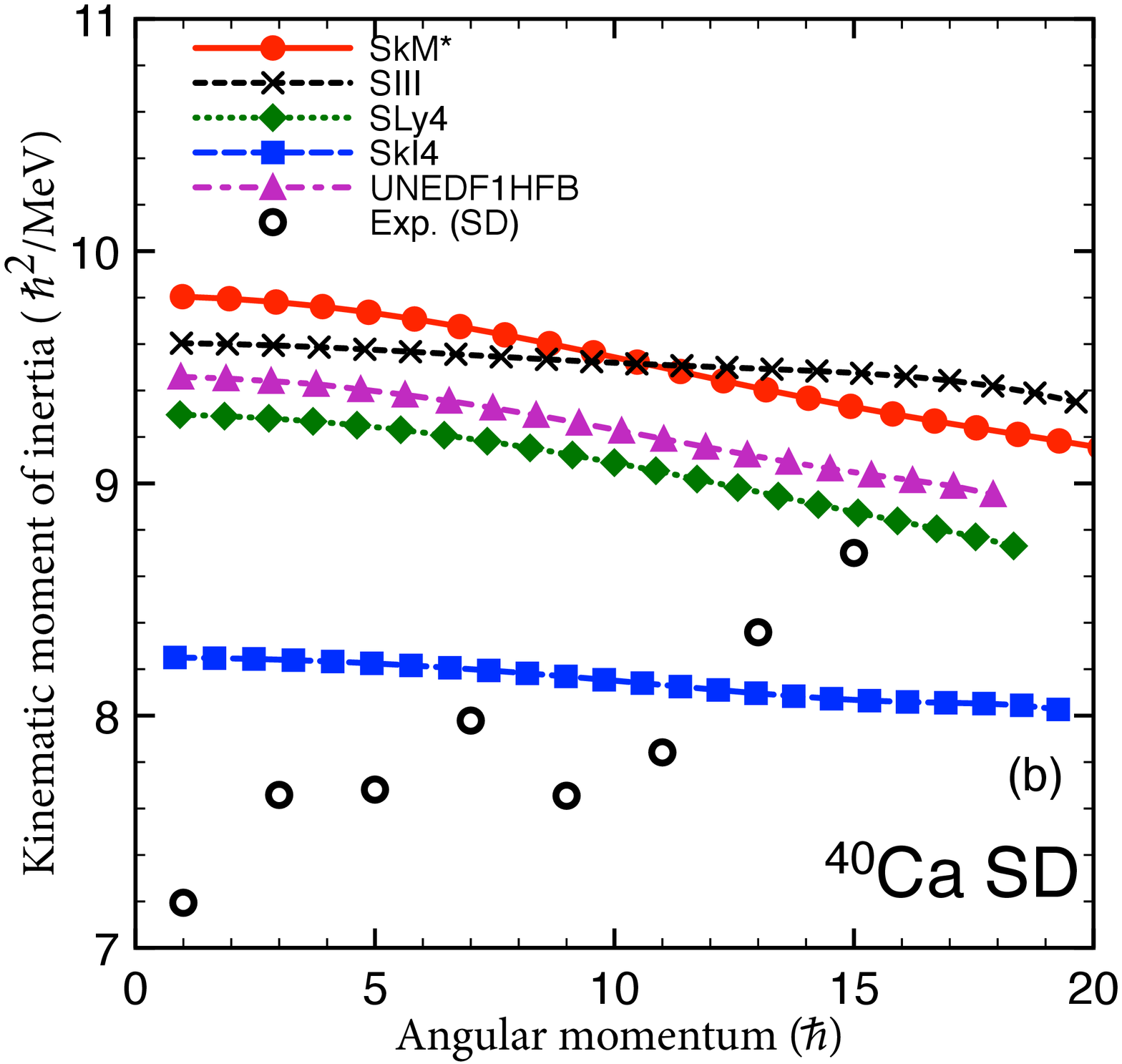}
\caption{
Excitation energies (a) and kinematic moments of inertia (b) of the reference SD rotational band 
in $^{40}$Ca as functions of angular momentum 
obtained by the cranked KS calculation employing the SkM*, SIII, SLy4, SkI4, and UNEDF1-HFB functionals 
together with the experimental data denoted as band 1 in Ref.~\cite{ide01}.
A smooth part $A I(I+1)$ is subtracted with an inertia parameter $A=0.05$ MeV for plotting the excitation energy.
}
\label{40Ca_SD}
\end{center}
\end{figure}

\begin{figure}[t]
\begin{center}
\includegraphics[scale=0.35]{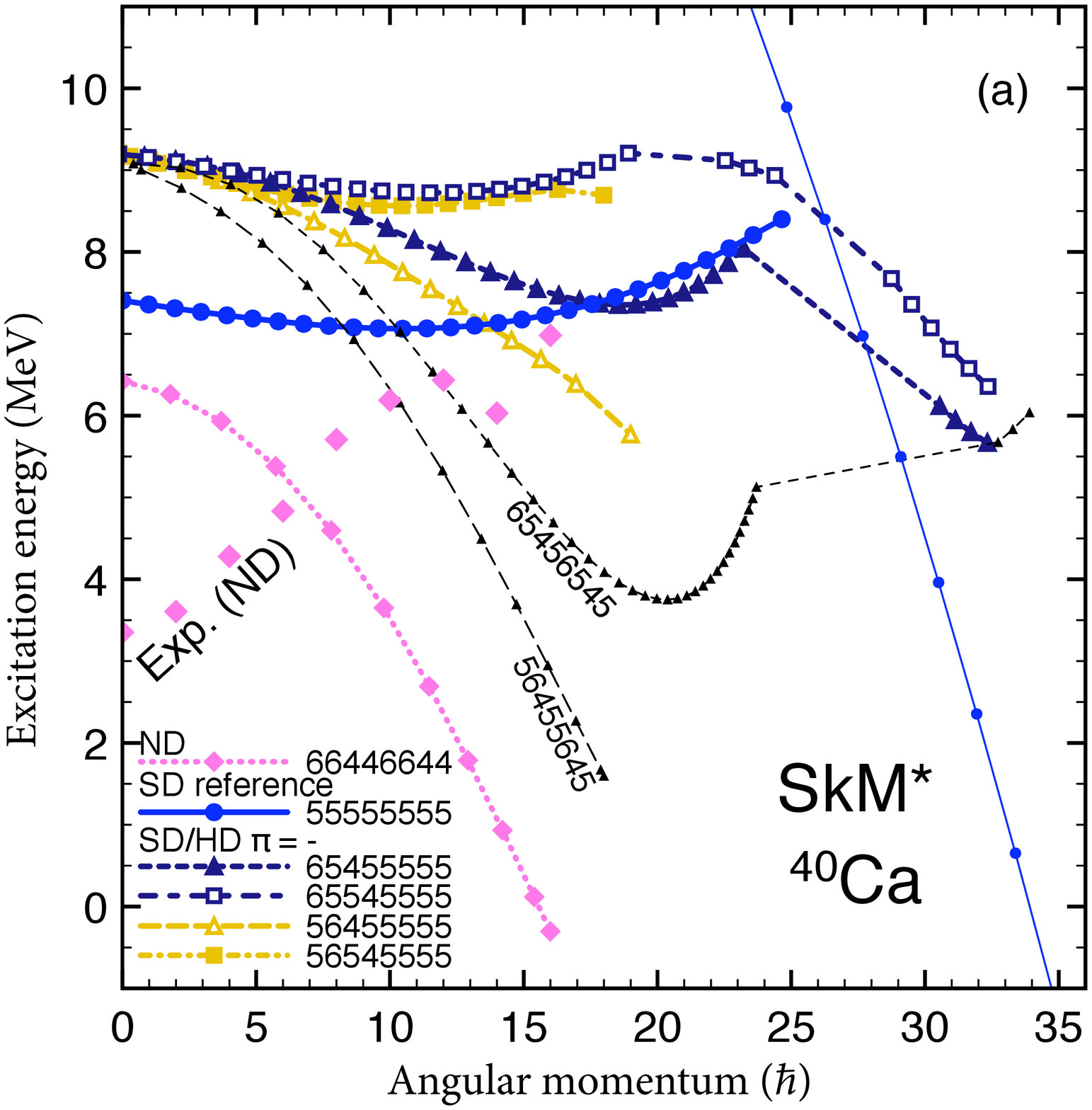}
\includegraphics[scale=0.35]{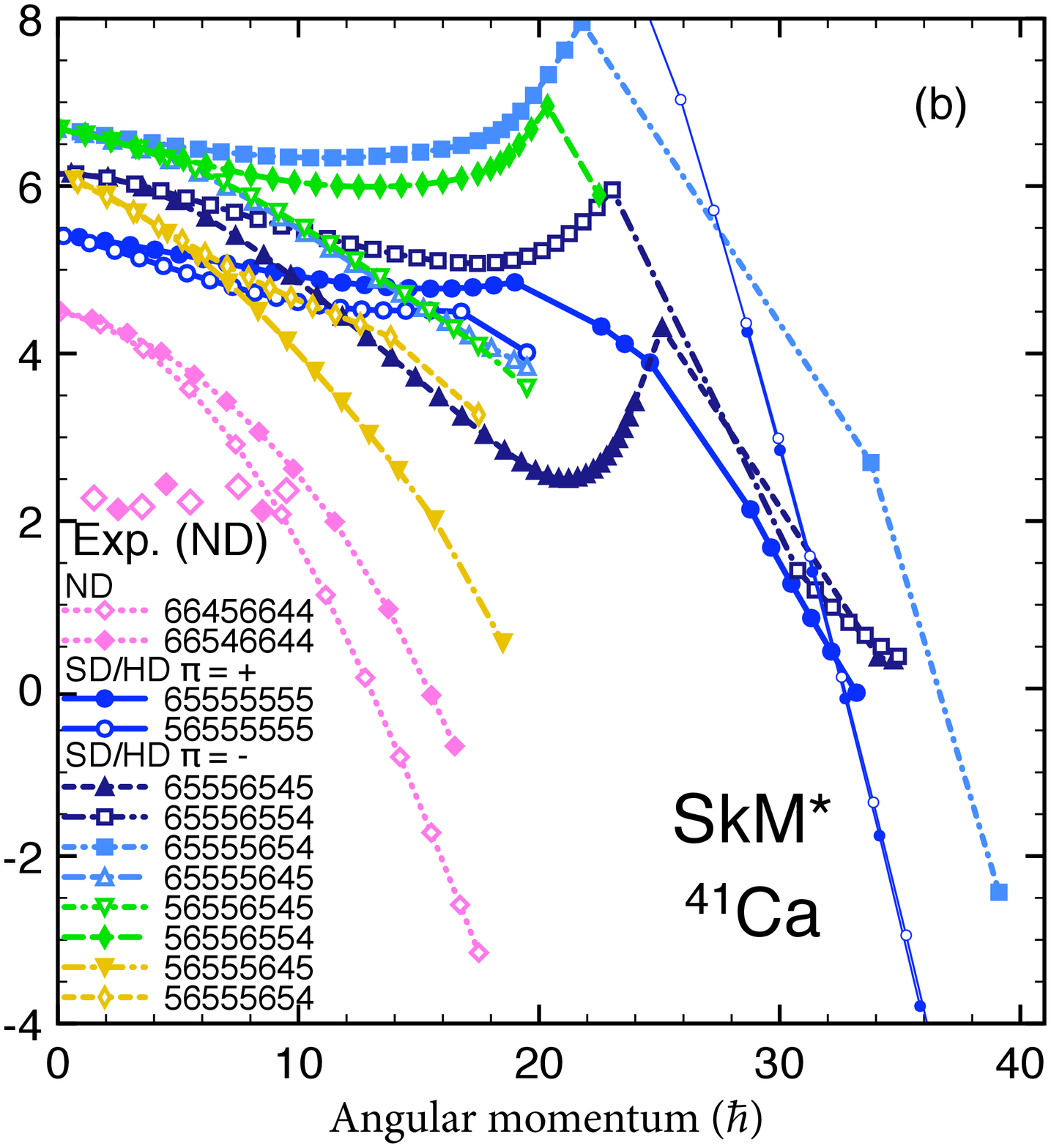}
\caption{
Excitation energies of the calculated configurations in $^{40}$Ca (a) and $^{41}$Ca (b) with the use of the SkM* functional. 
Filled (open) symbols indicate the signature exponent of the band is $\alpha=0 (1)$ and $+1/2 (-1/2)$ for $^{40}$Ca and $^{41}$Ca, respectively. 
A smooth part $A I(I+1)$ is subtracted with an inertia parameter $A=0.05$ MeV as in Fig.~\ref{40Ca_SD}. 
The experimental data for the normal deformation~\cite{ide01, bha16} are also shown. 
}
\label{Ca_band_SkM}
\end{center}
\end{figure}

\begin{figure}[t]
\begin{center}
\includegraphics[scale=0.33]{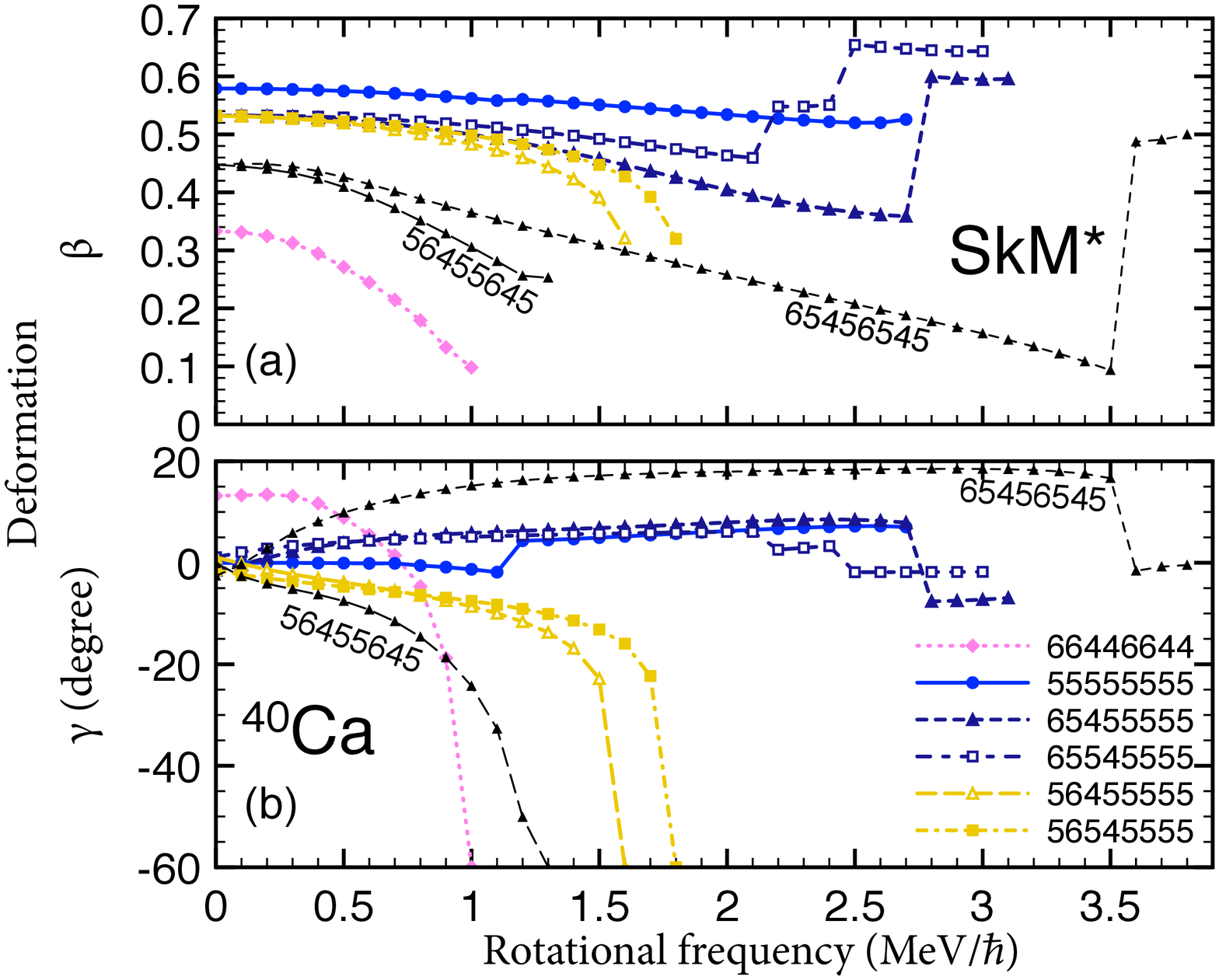}
\includegraphics[scale=0.33]{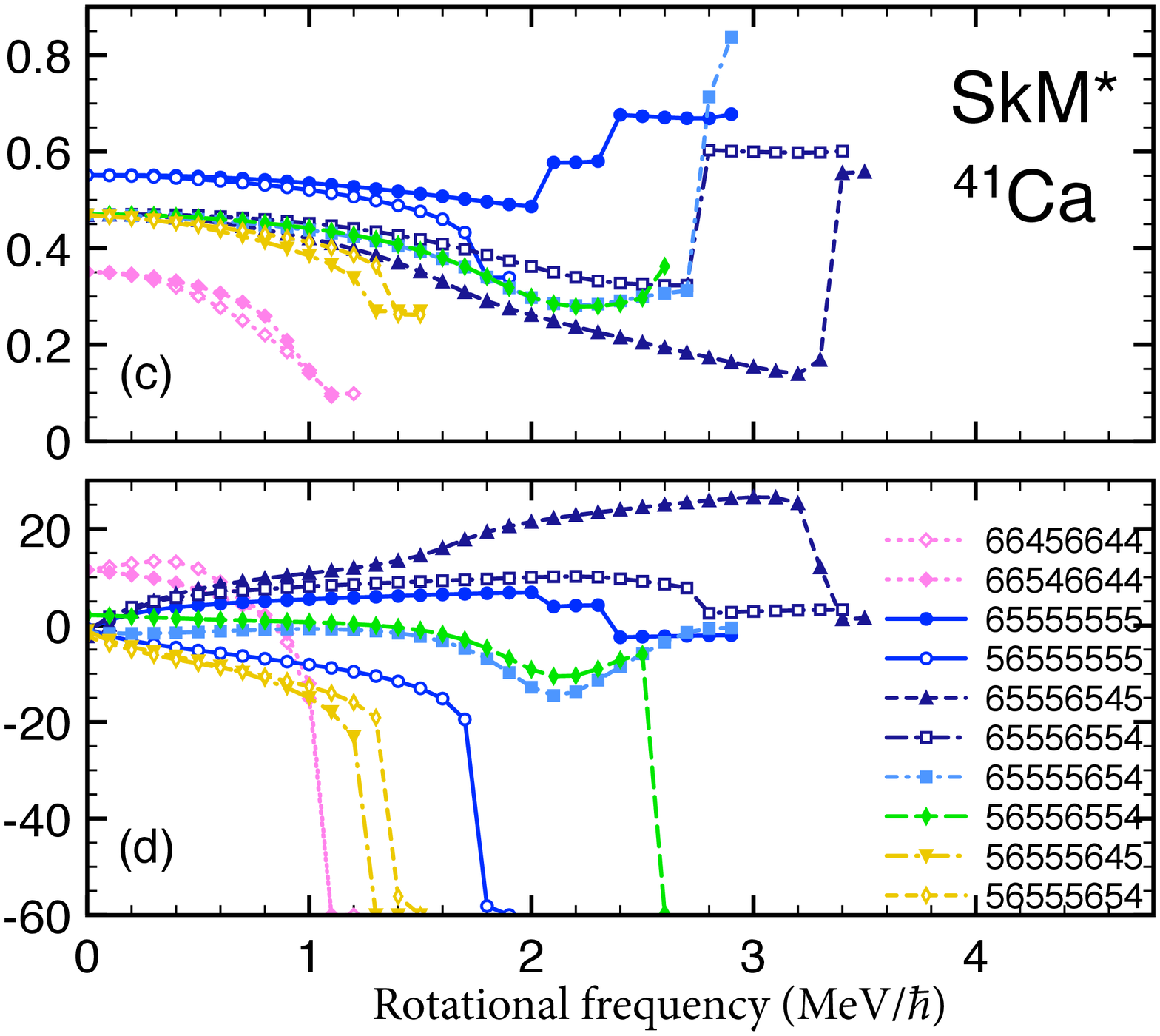}
\caption{
Evolution of quadrupole deformation, as a function of rotational frequency, 
for the calculated configurations in $^{40}$Ca [(a), (b)] and $^{41}$Ca [(c), (d)] with the use of the SkM* functional.
}
\label{Ca_def_SkM}
\end{center}
\end{figure}

Figure~\ref{40Ca_SD}(a) shows the calculated excitation energies, as functions of angular momentum, 
of the reference SD configuration of $^{40}$Ca obtained 
by employing the Skyrme EDFs including SkM*, SIII, SLy4, SkI4, and UNEDF1-HFB 
together with the experimental data~\cite{ide01}. 
The cranking calculations were carried out up to about $\omega_\mathrm{rot}=2.0$ MeV$/\hbar$ with an 
interval of $\Delta\omega_\mathrm{rot}=0.1$ MeV$/\hbar$. 
As usual, the angular momentum is evaluated as $I \hbar= \braket{J}_z$.
A smooth rigid-body part $A I(I+1)$ is subtracted with an inertia parameter $A=0.05$ MeV 
to make the difference of the results visible. 
The calculation with the Skyrme functionals tends to overestimate the observed excitation energy.  
And the calculated kinematic moments of inertia, $\mathcal{J}=I\hbar/\omega_{\rm rot}$, are large as shown in Fig~\ref{40Ca_SD}(b), 
which leads to a gentle slope in the energy vs. angular-momentum plot. 
Among the functionals we employed, 
the SkM* functional reproduces reasonably the observed excitation energy of the SD band in $^{40}$Ca. 
Thus we are going to discuss the high-spin states in $^{40,41}$Ca 
obtained mainly by using the SkM* functional. 
The SkI4 functional reproduces reasonably  
the observed kinematic moment of inertia as shown in Fig.~\ref{40Ca_SD}(b). 
Thus, the calculation employing the SkI4 functional will be used to complement the discussion.

Figures~\ref{Ca_band_SkM}(a) and \ref{Ca_def_SkM}(a), \ref{Ca_def_SkM}(b) show 
the excitation energies and the evolution of deformation of 
solutions with various intrinsic configurations in $^{40}$Ca obtained by the use of the SkM* functional. 
The low-spin spherical configuration $[7733]_\nu [7733]_\pi$ is not shown here because we are 
interested in the possible strongly-deformed structures at high spins. 
The triaxial ND configuration $[6644]_\nu [6644]_\pi$, 
corresponding to the configuration [2,2] calculated in Ref.~\cite{ray16}, 
terminates around $I=16$ with the structure $\nu(f_{7/2})^2(d_{3/2})^{-2} \otimes \pi(f_{7/2})^2(d_{3/2})^{-2}$, 
where the deformation reaches a weakly-deformed oblate shape as shown in Fig.~\ref{Ca_def_SkM}(b). 
Here, an oblate shape is indicated by $\gamma=\pm 60^\circ$, 
and in our definition of the triaxial deformation parameter $\gamma$ and the choice of rotation axis,  
$\gamma=-60^\circ$ represents a non-collective rotation, 
where the rotation axis is parallel to the symmetry axis. 
Thus, 
the sign of $\gamma$ in the present definition is different from that in  
the Lund convention~\cite{NR95} as well as in Ref.~\cite{ray16}. 
Furthermore, let us mention the notation of the configuration defined in Ref.~\cite{ray16}. 
The configuration is labeled by $[n_{N=3}(n_{N=4}), p_{N=3}(p_{N=4})]$, where 
$n_{N=3}$ and $n_{N=4}$ are the number of occupied neutrons in the $N=3$ intruder and $N=4$ hyperintruder shells, 
and 
$p_{N=3}$ and $p_{N=4}$ are the number of occupied protons in the $N=3$ and $N=4$ shells. 
When the $N=4$ shell is empty, $n_{N=4}$ and $p_{N=4}$ are omitted. 

One sees that 
the reference SD configuration $[5555]_\nu [5555]_\pi$ $(\pi=+1, r=+1)$ or the configuration [4,4] 
representing the four-neutron and four-proton excitation into the $pf$ shell 
appears as an SD-yrast band below $I \simeq 15$, 
the deformation of which is $\beta \simeq 0.55$ and $\gamma \simeq 3^\circ$. 
We were not able to trace the configuration $[5555]_\nu [5555]_\pi$ 
above $\omega_{\mathrm{rot}}=2.8$ MeV/$\hbar$ ($I \simeq 25$) consistently with the previous calculation~\cite{ina02} 
at which the quadrupole deformation is $\beta = 0.53$ and $\gamma = 7.1^\circ$.
Instead, we found another solution corresponding to the MD state [42,42] 
in the region $\omega_{\rm rot}=0$ -- 2.5 MeV$/\hbar$ indicated by the thin line in Fig.~\ref{Ca_band_SkM}(a), 
in which the two-neutron and two-proton are excited into the $g_{9/2}$ shell. 
Above $I \simeq 25$, the MD band appears as a yrast band as predicted in Refs.~\cite{ina02,ray16}. 
The quadrupole deformation is $\beta = 0.97, \gamma = -0.9^\circ$ at $\omega_{\rm rot}=2.5$ MeV$/\hbar$ ($I \simeq 39$).

Next, we explore possibility of strongly-deformed states with other configurations. As shown in 
Fig.~\ref{40Ca_SD_spe}, 
the single-particle orbitals just below and above the SD gap at $N, Z=20$ are $[321]3/2$ and
$[200]1/2$, respectively, for both neutrons and protons. 
As candidates which may appear near the yrast, 
we calculate all the possible ``1p1h'' and ``2p2h'' excitations from the reference SD  configuration 
associated with these two orbitals. 

There are four neutron ``1p1h'' excitations from $[321]3/2(r=\pm \ii)$ to 
$[200]1/2(r=\pm \ii)$, labeled as $[6545]_\nu, [5645]_\nu, [6554]_\nu$, and $[5654]_\nu$,
while keeping the proton configuration $[5555]_\pi$. 
They have negative parity 
and may be labeled also as [3,4] following Ref.~\cite{ray16}. 
Calculated results for these configurations
are shown in Figs.~\ref{Ca_band_SkM}(a), \ref{Ca_def_SkM}(a),  and \ref{Ca_def_SkM}(b). 
For all the four negative-parity configurations
we obtain solutions which commonly have large deformation $\beta\sim 0.5$ at low
spins $I \lesssim 12$, and hence can be regarded as SD bands. 
At higher spins, however, a difference grows up. 
Nevertheless we observe the following systematic trends.

The configuration $[5645]_\nu [5555]_\pi$ with quantum numbers $(\pi=-1, r=-1)$, i.e. 
a ph excitation of $\nu[321]3/2 (r=+\ii) \to \nu[200]1/2 (r=-\ii)$ across the SD gap, 
terminates around $I=20$. 
The configuration $[5654]_\nu [5555]_\pi$ with  $(\pi=-1, r=+1)$, 
another ph excitation of $\nu[321]3/2 (r=-\ii) \to \nu[200]1/2 (r=-\ii)$,
terminates also around $I=20$ similarly to $[5645]_\nu [5555]_\pi$. 
Both bands have negative-$\gamma$ deformation developing 
with an increase in the rotational frequency $\omega_{\rm rot}>0$, and reach 
$\gamma \simeq -60^\circ$ at the termination.
The other two negative-parity bands,
 $[6554]_\nu [5555]_\pi$ 
$(\pi=-1, r=-1)$ and  $[6545]_\nu [5555]_\pi$ $(\pi=-1, r=+1)$, in contrast, 
exhibit
rather stable deformation with relatively large $\beta$-deformation $\beta = 0.4$ -- $0.5$,
keeping the character of SD bands up to  
$ \omega_{\rm rot} \simeq 2.0$ MeV$/\hbar$ ($I \simeq 20$). It is also noticed that
these two bands have positive-$\gamma$, opposite sign of those of $[5645]_\nu [5555]_\pi$
and $[5654]_\nu [5555]_\pi$. 
These two bands have ph configurations from either  $\nu[321]3/2 (r=-\ii)$ or 
$\nu[321]3/2 (r=+\ii)$ to the common $\nu[200]1/2 (r=+\ii)$ orbital.
The above observations indicate that 
the occupation of the single-particle orbital $\nu[200]1/2 (r=+\ii)$
or $\nu[200]1/2 (r=-\ii)$ plays a decisive role not only to bring about the triaxial
deformation but also to determine the sign of $\gamma$:
 $\gamma<0$ in case $\nu[200]1/2 (r=-\ii)$ is occupied,
as in the first two configurations, $[5645]_\nu [5555]_\pi$ and
$[5654]_\nu [5555]_\pi$ , while $\gamma>0$ 
if  $\nu[200]1/2 (r=+\ii)$ is occupied
(the cases of $[6554]_\nu [5555]_\pi$  and  $[6545]_\nu [5555]_\pi$). 
Note that 
the positive and negative values for $\gamma$ deformation at $\omega_{\rm rot}>0$ imply that 
the energy-minimized state rotates about the intermediate and short axes, respectively, and that
a further increase of triaxiality with negative sign ($\gamma \rightarrow -60^\circ$) may lead to the
non-collective oblate rotation, i.e. the band termination. 

\begin{figure}[t]
\begin{center}
\includegraphics[scale=0.59]{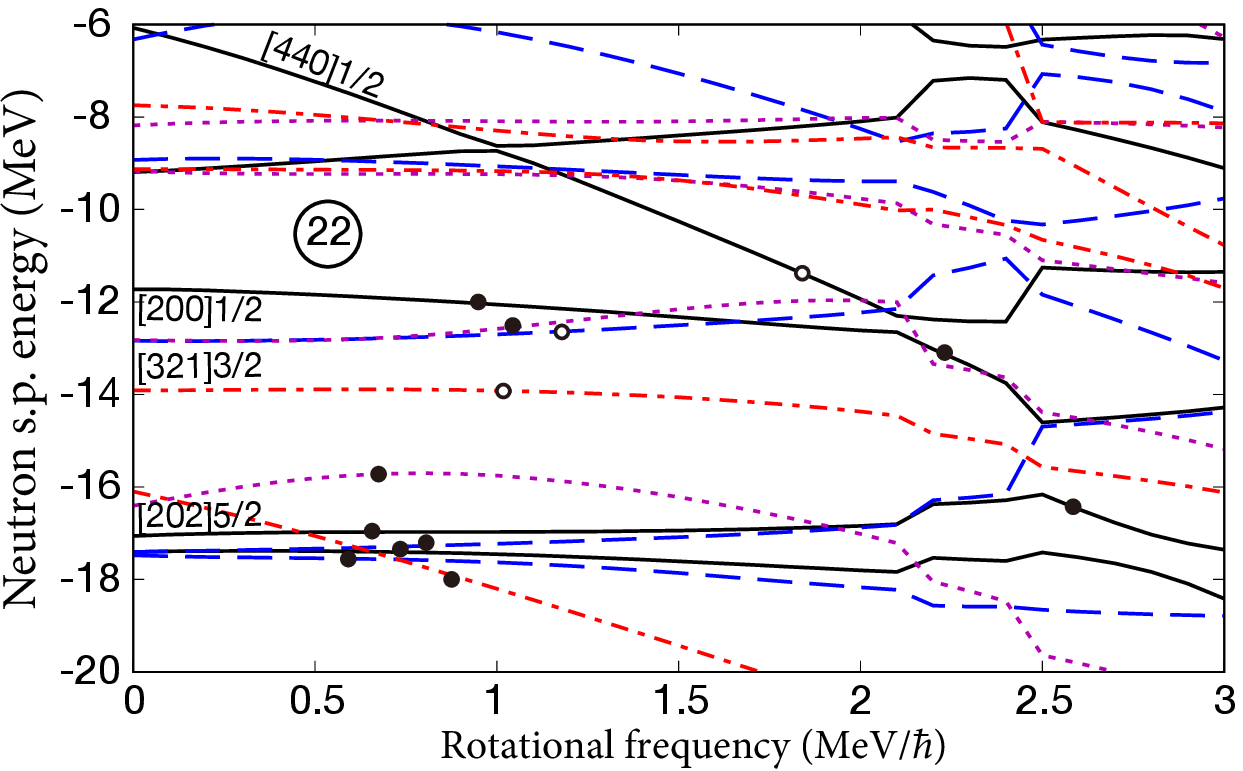}
\includegraphics[scale=0.59]{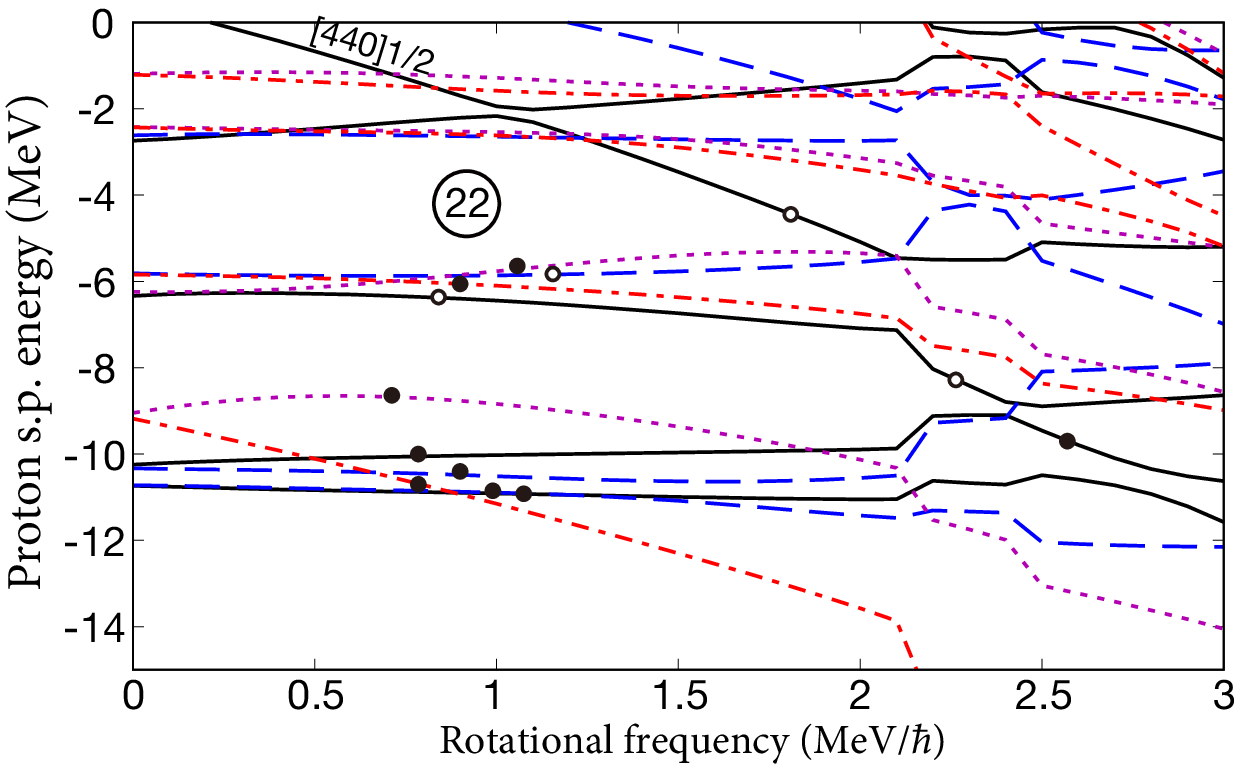}
\caption{
Neutron (left) and proton (right) single-particle energies (Routhians) in the 
selfconsistent rotating potential as functions of the rotational frequency. 
They are given for the $[6554]_\nu [5555]_\pi$ configuration ($\pi=-1, r=-1$) in $^{40}$Ca. 
Solid, long-dashsed, dashed, and dot-dashed lines indicate 
($\pi = +1, r=+\ii$), ($\pi = +1, r= -\ii$), ($\pi = -1, r = +\ii$), and ($\pi = -1, r=-\ii$) 
orbitals, respectively. 
Filled (open) circles indicate the orbitals occupied (empty). 
Some levels are labeled by the asymptotic quantum numbers $[N n_3 \Lambda]\Omega$ 
of the dominant component of the wave functions. 
}
\label{40Ca_6554}
\end{center}
\end{figure}

An interesting feature in $^{40}$Ca at high spin
is that the negative-parity SD bands with the configuration $[6554]_\nu [5555]_\pi$ 
$(\pi=-1, r=-1)$ and the configuration $[6545]_\nu [5555]_\pi$ $(\pi=-1, r=+1)$ 
are predicted to extend to even higher spins $I\gtrsim 20$ and 
appear near the yrast line beyond $I\simeq 25$. 
Furthermore, they exhibit stepwise increase
of deformation reaching $\beta \simeq 0.6$ -- 0.65. 
We show in Fig.~\ref{40Ca_6554} the rotational frequency dependence of the single-particle energy levels 
(Routhian) for the configuration $[6554]_\nu [5555]_\pi$. 
Due to the polarization associated with the time-odd mean field, 
the signature splitting is seen even at $\omega_\mathrm{rot}=0$. 
The $[200]1/2 (r=+\ii)$ orbital (solid line in Fig.~\ref{40Ca_6554}) 
is favored under rotation than the $[200]1/2 (r=-\ii)$ orbital (long-dashed line).
We see that 
the $\nu[440]1/2 (r=+\ii)$ orbital 
stemming from the $g_{9/2}$ orbital and the occupied $\nu[200]1/2 (r=+\ii)$ orbital 
cross in $I = 19$ -- $23$ ($\omega_\mathrm{rot}=2.2$ -- $2.3$ MeV$/\hbar$), 
resulting in the configuration [31,4]. 
And, the change of the proton configuration occurs in 
$I=24$ -- $28$ ($\omega_\mathrm{rot} = 2.4$ -- $2.5$ MeV$/\hbar$), where 
the $\pi[440]1/2 (r=+\ii)$ orbital crosses with the occupied $[202]5/2 (r=+\ii)$ orbital,   
resulting in a further development of deformation. 
These band crossings lead to the development of deformation from $\beta=0.45$ to 0.65 in total 
as shown in Fig.~\ref{Ca_def_SkM}(a) and bring in aligned angular momentum of $\sim 8$. 
Thus the configuration $[6554]_\nu [5555]_\pi$ at high spins corresponds 
to the ph excitation of $\nu[321]3/2 (r=-\ii$) $\to$ $\nu[440]1/2 (r=+\ii$) 
together with the ph excitation of $\pi[202]5/2 (r=+\ii)$ $\to$ $\pi[440]1/2 (r=+\ii)$ 
from the reference SD configuration $[5555]_\nu [5555]_\pi$.
Namely it is the HD configuration in which  the $[440]1/2$ orbitals of 
both neutron and proton are occupied, corresponding to the configuration [31,41]. 

Concerning the configuration $[6545]_\nu [5555]_\pi$, 
we found a successive occupation of the $[440]1/2 (r=+\ii)$ orbital 
of both neutron and proton in $I=23$ -- $31$
($\omega_\mathrm{rot}=2.7$ -- $2.8$ MeV$/\hbar$), 
resulting in the HD configuration where 
the ph excitation of $\nu[321]3/2 (r=+\ii) \to \nu[440]1/2 (r=+\ii)$ and 
the ph excitation of $\pi[202]5/2 (r=+\ii)$ $\to$ $\pi[440]1/2 (r=+\ii)$ are involved. 
The deformation increases suddenly from $\beta=0.35$ to $0.6$ 
and the spin increases with aligned angular momentum of $\sim 8$. 
Therefore, this state also corresponds to the HD configuration of [31,41] type.

It is emphasized here that the occupation of the $\nu [200]1/2 (r=+\ii)$ orbital 
in both configurations $[6554]_\nu [5555]_\pi$ and $[6545]_\nu [5555]_\pi$
have two roles: The first is to cause the positive-$\gamma$ deformation and
keep the large deformation up to high spins. The second
is to make the systems evolve from the SD to the HD configurations through the avoided level crossing 
between the $\nu [200]1/2 (r=+\ii)$ and the  $[440]1/2 (r=+\ii)$ orbitals.
On the other hand, the occupation of the signature-partner $\nu [200]1/2 (r=-\ii)$ orbital 
favors the negative-$\gamma$ deformation at $\omega_{\rm rot}>0$ then the 
configurations occupying this orbital 
terminate around $I=20$ leading to a weakly deformed oblate shape.
Therefore, the occupation of the $[200]1/2 (r=+\ii)$ orbital at low spins is necessary 
for the shape evolution towards the HD bands with an increase in spin.

We here remark relation to the preceding cranked EDF studies.
It is predicted in Ref.~\cite{ray16} that negative-parity excited bands with
neutron ``1p1h'' configuration [3,4] as well as proton ``1p1h'' configuration [4,3] 
have appreciable triaxiality with positive $\gamma \sim 10^\circ$
(negative $\gamma \sim -10^\circ$ in the convention adopted in Ref.~\cite{ray16}),
however the sign and the development of triaxiality are not deeply investigated.
We note also that in the present study we obtained proton ph-excited states [4,3], i.e,
$[5555]_\nu [6554]_\pi$ etc., which appear closely to the 
neutron ph-excited states because the system under consideration is a light $N=Z$ nucleus 
though we do not show these states in the present article. 
The random-phase approximation (RPA) in the rotating mean field is more appropriate
to describe the excited bands~\cite{mar76,jan79,egi80a,egi80b,zel80,shi83,shi84}, 
however it is beyond the scope of the present work.
The low-lying $K^\pi=1^-$ state on the SD $I^\pi=0^+$ state predicted in an RPA description~\cite{yos05,ina06} 
is weakly collective, and is constructed predominantly by the ph excitation $[321]3/2 \to [200]1/2$ 
of neutron and proton. 
This state can be regarded approximately as the superposition of the [3,4] and [4,3] configurations. 

Next we consider ``2p2h'' excitations with respect to the reference SD configuration, which have 
``1p1h'' excitations $[321]3/2 \to [200]1/2$  both in neutrons and protons and hence positive parity.
Some of them appear actually 
as yrast bands at the intermediate spins. 
We show in Fig.~\ref{Ca_band_SkM}(a) such examples; 
the configurations $[6545]_\nu [6545]_\pi$ and $[5645]_\nu [5645]_\pi$, 
corresponding to the 
configurations [3,3] calculated in Ref.~\cite{ray16}.  
The deformation of these states is not very large and 
the deformation parameter is around $\beta \sim 0.4$, 
which is in between the deformations of ND and SD configurations. 
This is because a ``2p2h'' excitation on the reference SD configuration can be also regarded as 
a ``2p2h'' excitation on the ND configuration. 
We found that the triaxial deformation develops with positive- and negative-$\gamma$ 
for the configurations $[6545]_\nu [6545]_\pi$ and $[5645]_\nu [5645]_\pi$, respectively 
[see Fig.~\ref{Ca_def_SkM}(b)].
The latter terminates quickly at $I \simeq 18$. 
The former configuration develops to the HD state with a band crossing 
where both a neutron and a proton are promoted to the $[440]1/2 (r=+\ii)$ orbital, 
corresponding to the configuration [31,31] representing 
the one-neutron and one-proton excitation into the $g_{9/2}$ shell.  
These behaviors can be explained by the mechanism discussed above, i.e. 
the occupation of the $[200]1/2 (r=+\ii)$ or $[200]1/2 (r=-\ii)$
orbital favoring the positive- and negative-$\gamma$ deformations, respectively, 
as increasing the rotational frequency.

\subsection{Superdeformation and hyperdeformation in $^{41}$Ca: role of the [200]1/2 orbital  \label{res_41Ca}}

\begin{table}[b]
\caption{Calculated configurations for $^{41}$Ca}
\label{41Ca_config}
\centering
\scalebox{0.84}[0.84]{
    \begin{tabular}{ccccc}
      \hline
     $\pi$  & $r$ & 5p5h+$n$ (6p5h) & 6p6h+$n$ (7p6h) & 8p8h+$n$ (9p8h) \\ \hline \hline
      $+1$ & $+\ii$ & $[6654]_\nu[6545]_\pi,[6654]_\nu[5654]_\pi,[6645]_\nu[5645]_\pi,[6645]_\nu[6554]_\pi$ & $[6555]_\nu[6644]_\pi$ & $[6555]_\nu[5555]_\pi$ \\ \cline{2-5}
       & $-\ii$ & $[6654]_\nu[5645]_\pi,[6654]_\nu[6554]_\pi,[6645]_\nu[6545]_\pi,[6645]_\nu[5654]_\pi$ & $[5655]_\nu[6644]_\pi$ & $[5655]_\nu[5555]_\pi$ \\ \hline
       \multicolumn{2}{c}{} & 7p7h+$n$ (8p7h) & 6p6h+$n$ (7p6h) & 4p4h+$n$ (5p4h) \\ \hline \hline
      $-1$ & $+\ii$ & $[6555]_\nu[6545]_\pi,[6555]_\nu[5654]_\pi,[5655]_\nu[5645]_\pi,[5655]_\nu[6554]_\pi$ & $[6654]_\nu[5555]_\pi$ & $[6654]_\nu[6644]_\pi$  \\ \cline{2-5}
       & $-\ii$ & $[6555]_\nu[5645]_\pi,[6555]_\nu[6554]_\pi,[5655]_\nu[6545]_\pi,[5655]_\nu[5654]_\pi$ & $[6645]_\nu[5555]_\pi$ & $[6645]_\nu[6644]_\pi$   \\ \hline
    \end{tabular}}
\end{table}

For the occurrence of the HD states at high spins in $^{40}$Ca, we found 
that the occupation of the $[200]1/2 (r=+\ii)$ orbital is necessary 
because the configurations involving 
this orbital favor the positive-$\gamma$ deformation at $\omega_{\mathrm{rot}}>0$, 
while the configurations involving the $[200]1/2 (r=-\ii)$ orbital favor the negative-$\gamma$ 
deformation at $\omega_{\rm rot}>0$ and terminate shortly. 
In what follows, 
we are going to investigate the possible appearance of the HD states
in $^{41}$Ca with paying attention to a role of the $[200]1/2$ orbital.

Figure~\ref{Ca_band_SkM}(b) shows the excitation energies for 
the near-yrast configurations of $^{41}$Ca calculated 
with the use of the SkM* functional.  
Actually, we tried 24 configurations, as listed in Table~\ref{41Ca_config}. 
Among positive parity configurations, we focus on the ones
$[6555]_\nu [5555]_\pi$ and $[5655]_\nu [5555]_\pi$ in which the last neutron
is added in the $\nu[200]1/2 (r=\pm \ii)$ orbitals on top of the reference SD configuration
$[5555]_\nu [5555]_\pi$ (denoted also 8p8h+$n$ in Table~\ref{41Ca_config}). For
negative parity configurations we shall discuss those with the last neutron in
$\nu[200]1/2 (r=\pm \ii)$ combined with proton ph-excitations $[6545]_\pi$ etc.
from the reference SD configuration, denoted 7p7h+$n$. 
We shall discuss also
$[6645]_\nu [6644]_\pi$ and $[6654]_\nu [6644]_\pi$, 
in which the last neutron in $\nu[321]3/2 (r=\pm\ii)$ is added to the ND configuration 
$[6644]_\nu [6644]_\pi$ in $^{40}$Ca,  denoted 4p4h+$n$. 
We show in Figs.~\ref{Ca_def_SkM}(c) and \ref{Ca_def_SkM}(d) the evolution of deformation for 
these configurations.

The triaxial-ND configuration $[6645]_\nu [6644]_\pi$ $(\pi=-1, r=-\ii)$
and its signature partner $[6654]_\nu [6644]_\pi$ $(\pi=-1, r=+\ii)$ 
appear as the yrast states and 
terminate around $I=18$ similarly as in the case of $^{40}$Ca. 
Experimentally, the $K^\pi=3/2^-$ band is reported up to $I=19/2$~\cite{bha16}, and 
it is shown in Fig.~\ref{Ca_band_SkM}(b) though  
the pairing correlations become important in such low spins 
and a direct comparison cannot be made.

Let us look at the positive-parity 
configuration [6555]$_{\nu}$ [5555]$_{\pi}$ $(\pi=+1, r=+\ii)$ corresponding to the configuration [4,4]. 
At $\omega_\mathrm{rot}=0$, this configuration 
is nothing but the one neutron added in the $[200]1/2 (r=+\ii)$ orbital to the reference SD configuration 
$[5555]_\nu [5555]_\pi$ in $^{40}$Ca. 
In $I=19$ -- $23$ ($\omega_\mathrm{rot}=2.0$ -- $2.1$ MeV$/\hbar$) and in 
$I=25$ -- $29$ ($\omega_\mathrm{rot}=2.3$ -- $2.4$ MeV$/\hbar$), one sees a successive band crossing: 
The occupied $\nu[200]1/2 (r=+\ii)$ orbital crosses with the $\nu[440]1/2 (r=+\ii)$ orbital, 
namely it corresponds to the configuration [41,4], then 
the deformation increases from $\beta=0.5$ to $0.6$ and the spin increases by $\sim 4$. 
And, the occupied $\pi[202]5/2 (r=+\ii)$ orbital crosses with the $\pi[440]1/2 (r=+\ii)$ orbital 
causing a further development of deformation to reach $\beta = 0.7$ with an additional aligned angular momentum of $\sim 4$
whose structure corresponds to the configuration [41,41]. 
Both a neutron and a proton occupy the $g_{9/2}$ shell. 
On the other hand, its signature partner, the configuration $[5655]_\nu [5555]_\pi$ $(\pi=+1, r=-\ii)$
also corresponding to the configuration [4,4], 
terminates around $I=20$. This is because the occupation of the $\nu [200]1/2 (r=-\ii)$ orbital 
leads to the negative-$\gamma$ deformation at $\omega_{\rm rot}>0$.
The signature-partner SD bands reveal an opposite evolution of triaxiality 
depending on the signature quantum number of a neutron which is added to the reference SD state in $^{40}$Ca.  
Note that we found the solutions corresponding to the MD configuration as shown by the thin lines in Fig.~\ref{Ca_band_SkM}(b), 
and the signature-partner MD bands may appear as the yrast bands above $I \simeq 34$.

Then, we are going to discuss the excited negative-parity bands. 
The configuration [6555]$_{\nu}$ [6545]$_{\pi}$ with quantum numbers$(\pi=-1, r=+\ii)$, 
denoted by filled triangle in Fig.~\ref{Ca_band_SkM}(b), 
appears as a yrast band for $20 \lesssim I \lesssim 26$. 
At $\omega_\mathrm{rot}=0$, this configuration 
gives a large deformation with $\beta = 0.47$, 
and corresponds to 
the ph excitation of $\pi[321]3/2 (r=+\ii) \to \pi[200]1/2 (r=+\ii)$ across the SD gap. 
With an increase in the angular momentum, the deformation is getting smaller: 
The deformation is only $\beta \sim 0.3$ but the triaxiality is large with $\gamma \simeq 18^\circ$ at $I \simeq 18$. 
In $I=25$ -- 34, we see a band crossing, 
where both the neutron and proton $[200]1/2 (r=+\ii)$ orbitals cross with 
the $[440]1/2 (r=+\ii)$ orbitals, resulting in the HD configuration, i.e. [41,31] type, and 
bringing in aligned angular momentum of $\sim 9$.  

\begin{figure}[t]
\begin{center}
\includegraphics[scale=0.58]{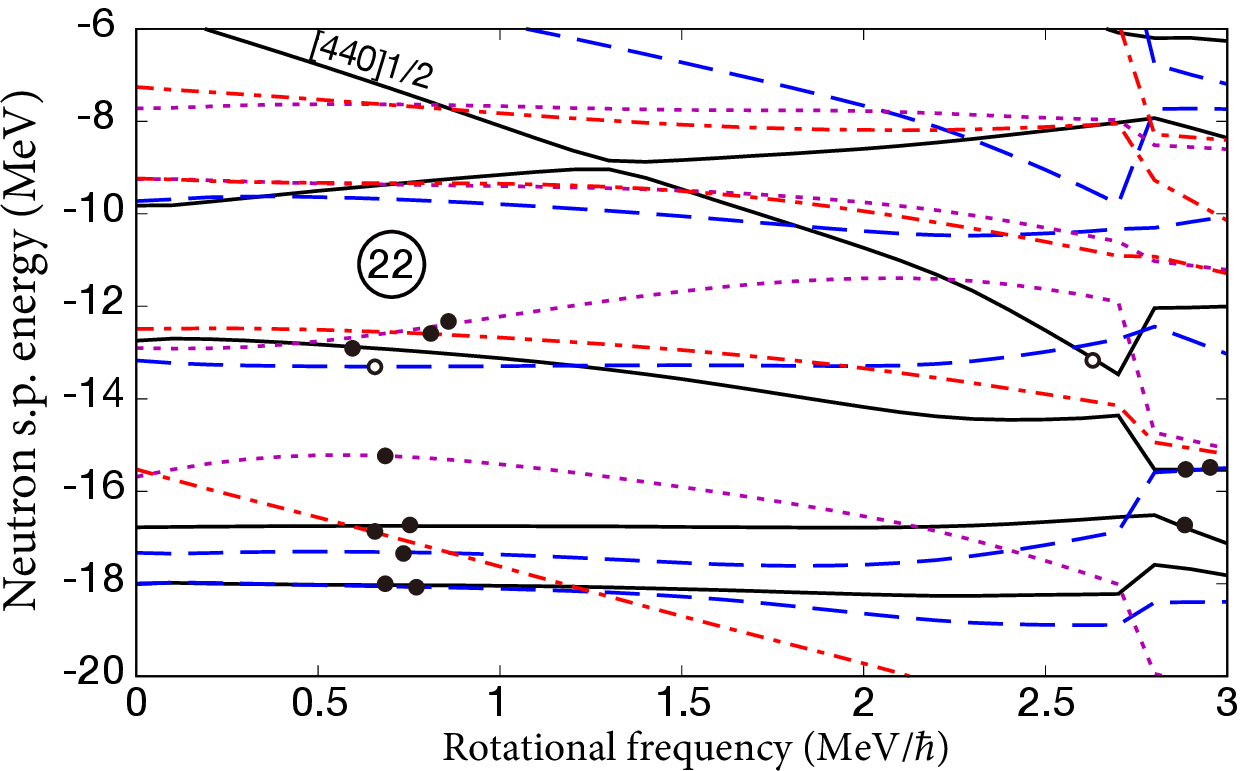}
\includegraphics[scale=0.58]{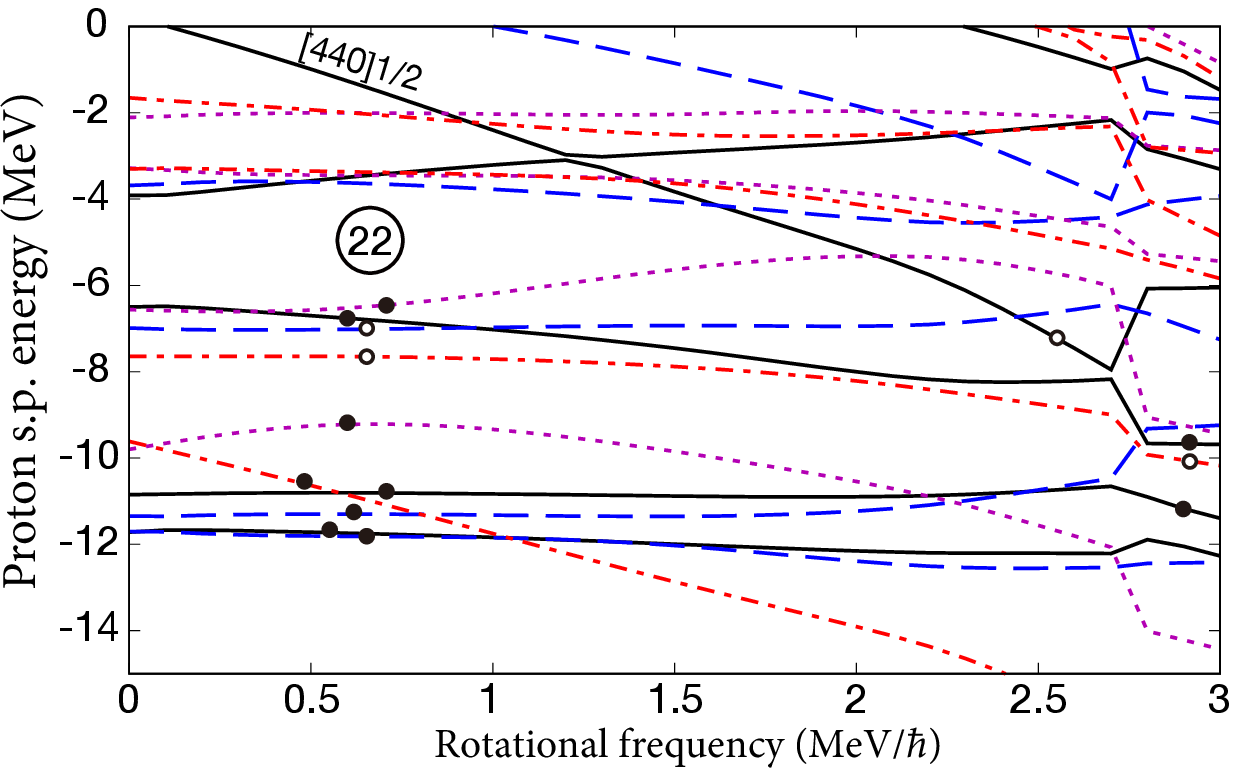}
\caption{
Same as Fig.~\ref{40Ca_6554} but for the configuration of $[6555]_\nu [6554]_\pi$ ($\pi=-1, r=-\ii$)
in $^{41}$Ca. 
}
\label{41Ca_6555}
\end{center}
\end{figure}

The configuration $[6555]_\nu [6554]_\pi$ with quantum numbers$(\pi=-1, r=-\ii)$, 
denoted by open square in Fig.~\ref{Ca_band_SkM}(b),  
appears closely to the yrast band at high spins $I \gtrsim 30$. 
Figure~\ref{41Ca_6555} shows the Routhians for this configuration.
Both a neuron and a proton occupy the $[200]1/2 (r=+\ii)$ orbital at $\omega_\mathrm{rot}=0$. 
With an increase in the rotational frequency, 
the occupied $[200]1/2 (r=+\ii)$ orbital of both neutron and proton cross with the $[440]1/2 (r=+\ii)$ orbital 
in $I = 23$ -- $31$ ($\omega_\mathrm{rot}=2.7$ -- $2.8$ MeV$/\hbar$). 
Therefore, one sees a sudden increase of deformation at $\omega_\mathrm{rot} \simeq 2.7$ MeV$/\hbar$ 
from $\beta=0.3$ to $0.6$ with aligned angular momentum of $\sim 8$. 
This state corresponds to the configuration [41,31] and 
is regarded as a family of the HD band of $[6554]_\nu [5555]_\pi$ in $^{40}$Ca.

These examples indicate 
that the occupation of the $[200]1/2 (r=+\ii)$ orbital at low spins 
plays an important role for the appearance of the HD state at high spins. 
This can be explained in terms of the $[200]1/2(r=\pm\ii)$ orbitals, 
which we have found in the SD bands in $^{40}$Ca. 
When both a neutron and a proton occupy the $[200]1/2 (r=+\ii)$ orbital at low spins, 
we obtain the HD configuration at high spins thanks to the level crossing with the $[440]1/2(r=+\ii)$ orbital, 
while the band terminates shortly after increasing the spin for the case that 
both a neutron and a proton occupy the $[200]1/2 (r=-\ii)$ orbital. 
This is because the occupation of the $[200]1/2 (r=+ \ii)$ and $[200]1/2 (r=-\ii)$ 
orbital leads to the positive-  and negative-$\gamma$ 
deformation at $\omega_{\rm rot} >0$, respectively. 
When both the $[200]1/2 (r=+\ii)$ and $[200]1/2 (r=-\ii)$ orbitals are occupied 
by a neutron and a proton, we have a competitive behavior for the development of triaxial deformation 
as seen for the $[6555]_\nu [5654]_\pi$ and $[5655]_\nu [6554]_\pi$ configurations.

\subsection{Case for the SkI4 functional  \label{SkI}}

One may doubt the above finding can be a specific feature of the EDF employed. 
We try to dispel the suspicion by investigating the functional dependence of the high-spin structures 
in $^{40}$Ca and $^{41}$Ca. 
We employ here the SkI4 functional as mentioned previously. 
As one can see in Figs.~\ref{Ca_def_SkI}(b) and \ref{Ca_def_SkI}(d), the SkI4 functional 
gives non-zero triaxiality at $\omega_{\rm rot}=0$ for most of the configurations under study.
Therefore, we expect  to see clearly 
the role of signature-dependent triaxiality in shape evolution of the SD states.

Before discussion, let us mention how to choose the sign of $\gamma$ at $\omega_{\rm rot}=0$.  
At $\omega_{\rm rot}=0$, the sign of $\gamma$ gives no difference, which means that 
we have two minima, with the same energy, in the potential energy surface 
in terms of the quadrupole deformations in 
the region of $-60^\circ < \gamma < 120^\circ$. 
At $\omega_{\rm rot}>0$, the asymmetry shows up in the total Routhian surface 
and we can choose the configuration that produces the lowest energy. 
We thus chose the sign of $\gamma$ at $\omega_{\rm rot}=0$, 
whose configuration gives lower energy at $\omega_{\rm rot}>0$ continuously.

\begin{figure}[t]
\begin{center}
\includegraphics[scale=0.35]{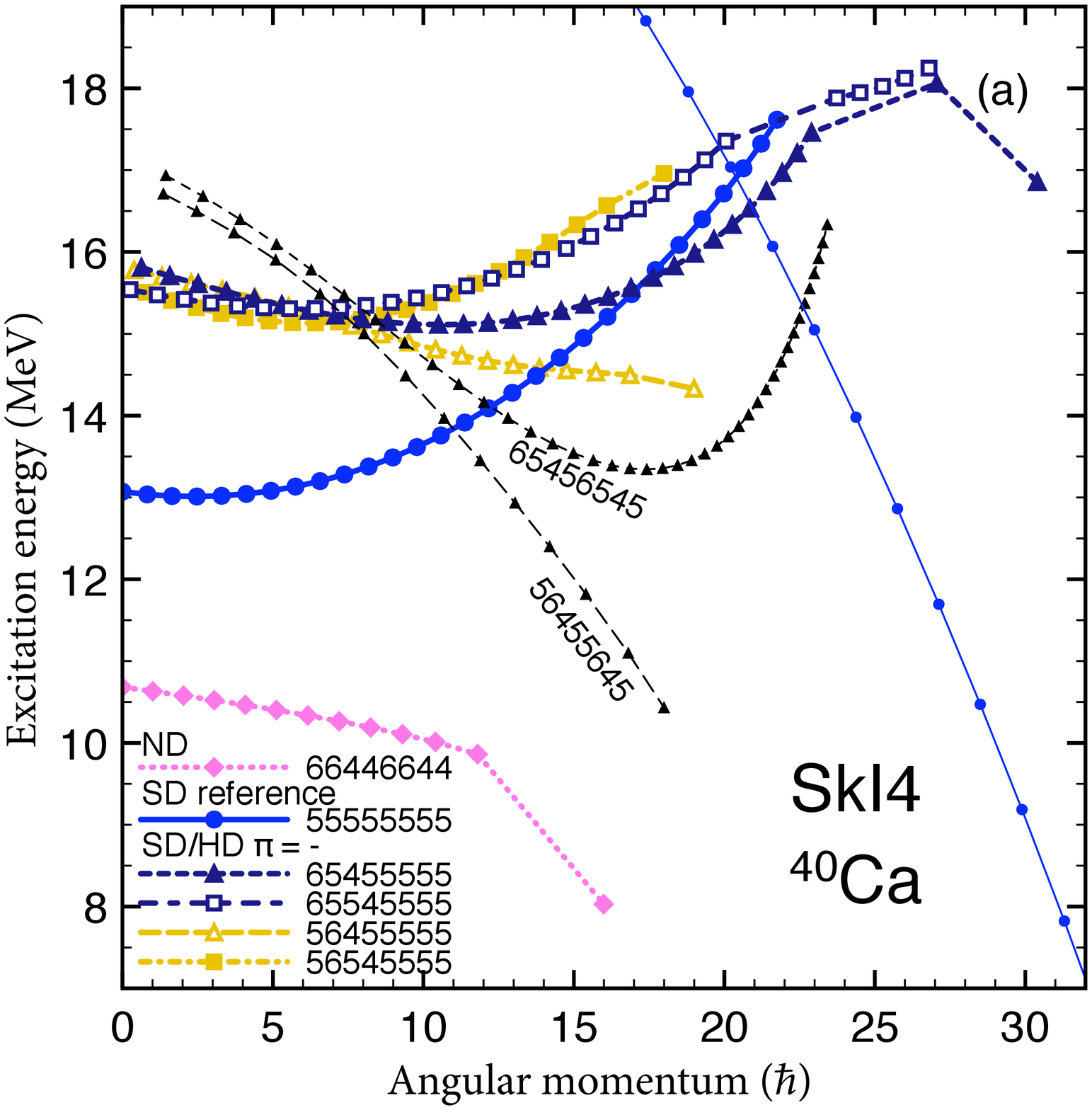}
\includegraphics[scale=0.35]{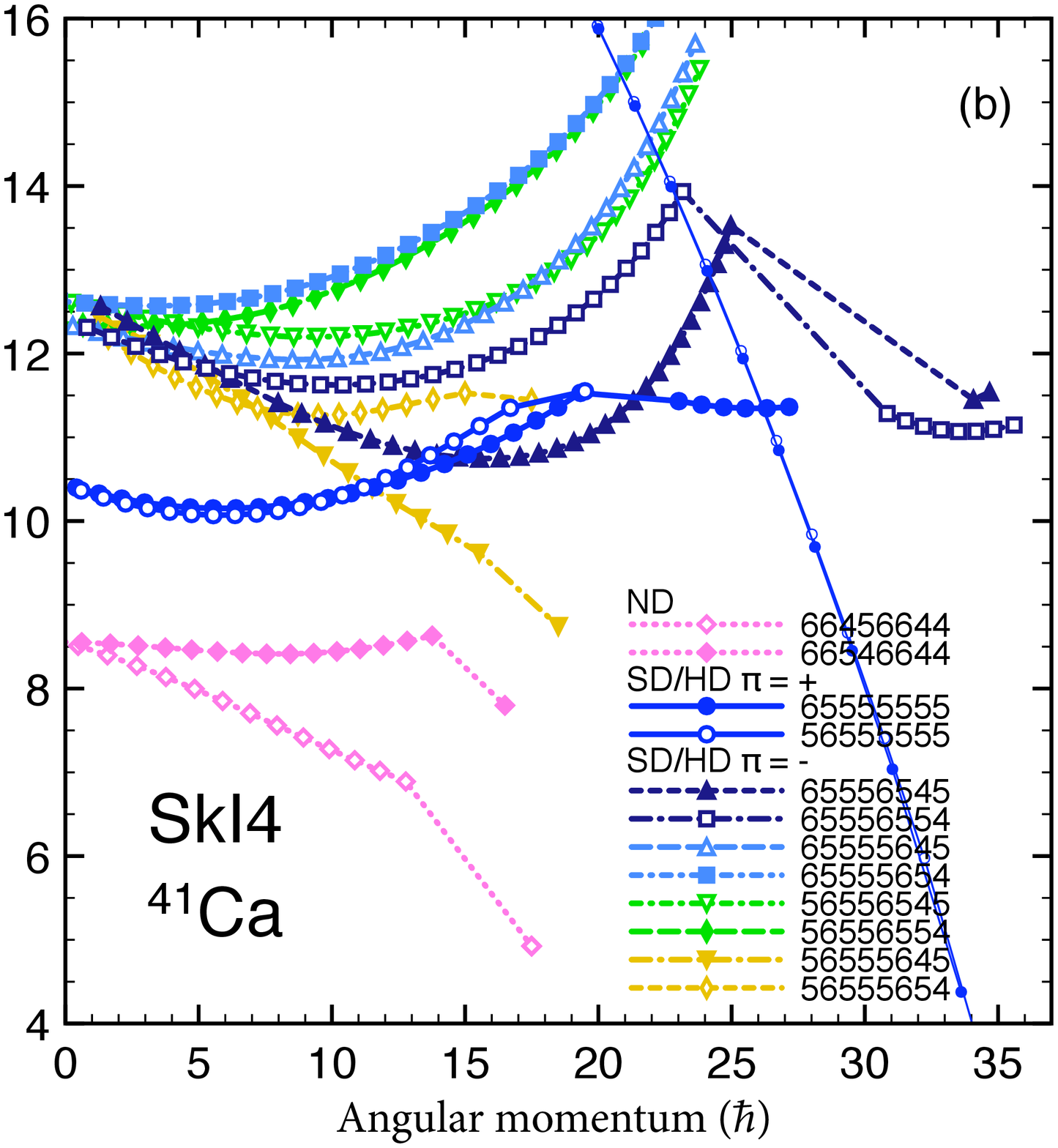}
\caption{
Same as Fig.~\ref{Ca_band_SkM} but obtained with the use of the SkI4 functional. 
}
\label{Ca_band_SkI}
\end{center}
\end{figure}

\begin{figure}[t]
\begin{center}
\includegraphics[scale=0.33]{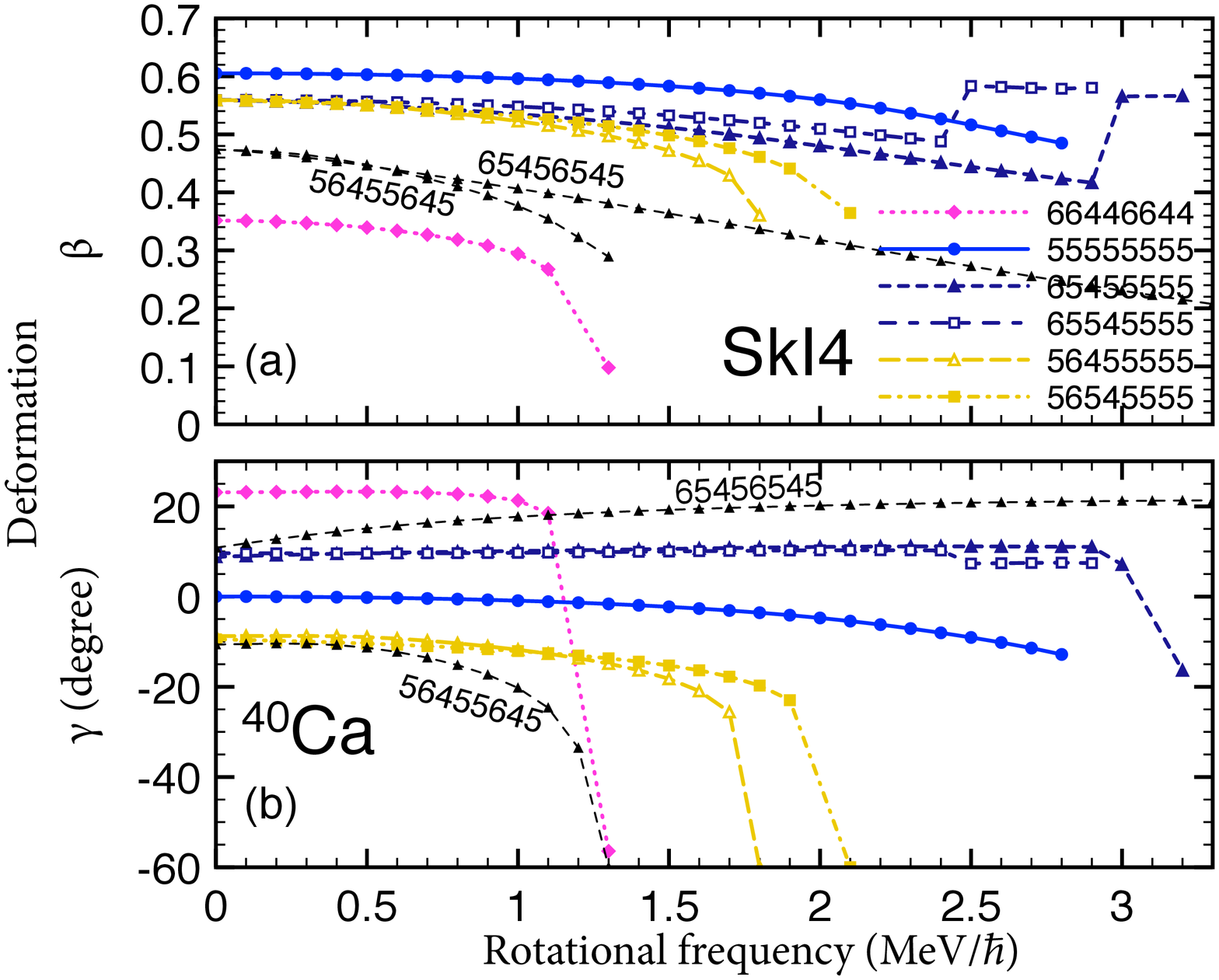}
\includegraphics[scale=0.33]{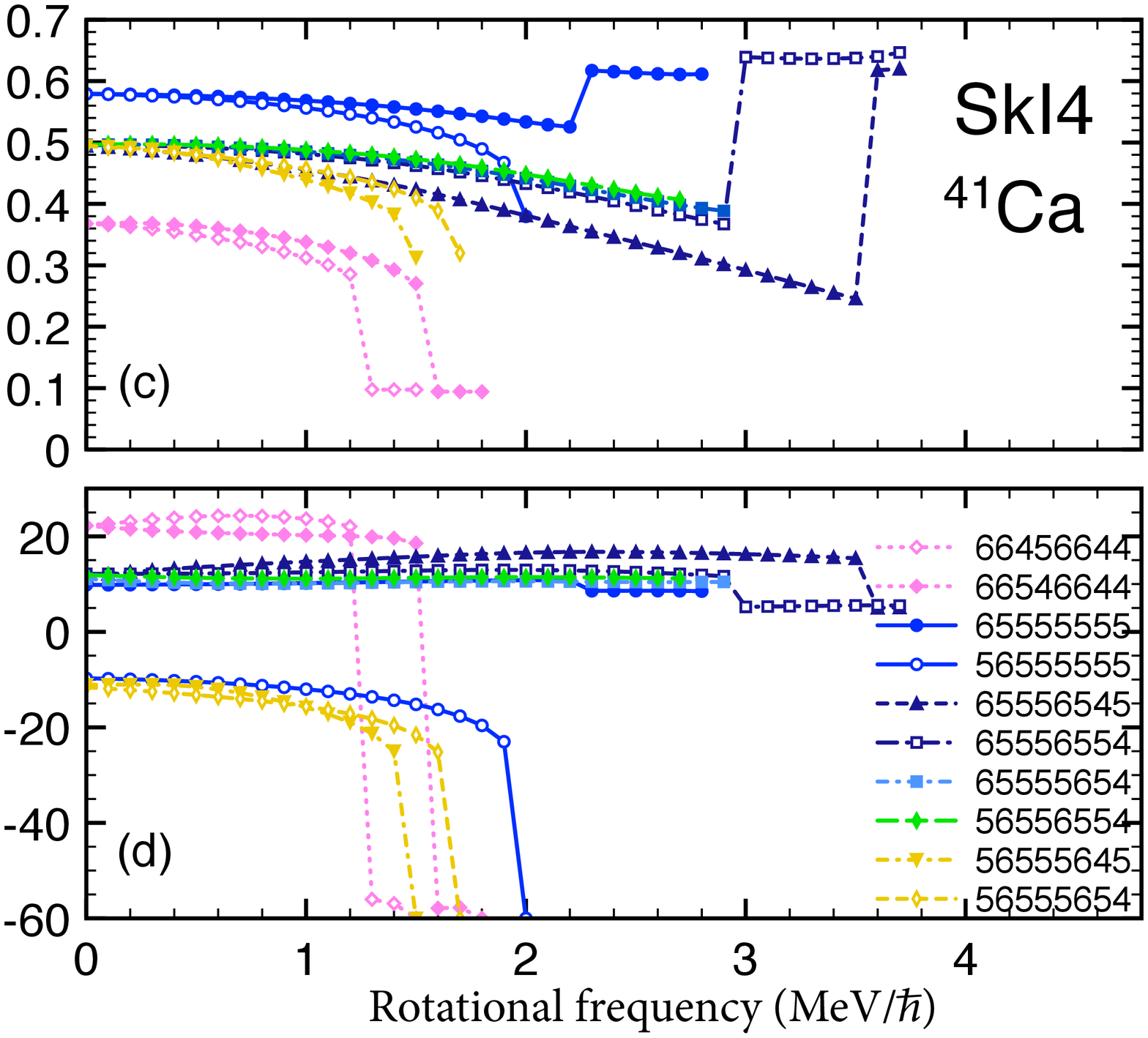}
\caption{
Same as Fig.~\ref{Ca_def_SkM} but obtained with the use of the SkI4 functional.
}
\label{Ca_def_SkI}
\end{center}
\end{figure}

Then, we are going to investigate the near-yrast structures in $^{40}$Ca at first. 
Figures~\ref{Ca_band_SkI}(a) and \ref{Ca_def_SkI}(a), \ref{Ca_def_SkI}(b) 
show the excitation energies and deformation. 
It is seen from comparison of the excitation energies 
[Fig.~\ref{Ca_band_SkM}(a) vs. Fig.~\ref{Ca_band_SkI}(a)] 
that the relative ordering 
of the near-yrast SD bands and the ND band are essentially the same as that for SkM*.
On the other hand, a difference from SkM* is seen in the absolute magnitude of the excitation energies. 
For example, the negative-parity excited SD bands are well separated from the reference SD band
at low spins 
since the SD gap energy at the particle number 20 is higher than that calculated with SkM*.
This is similar to the results obtained by the CRMF calculation~\cite{ray16}.

The deformation of the obtained SD and ND states in $^{40}$Ca, 
shown in Figs.~\ref{Ca_def_SkI}(a) and \ref{Ca_def_SkI}(b), is 
also consistent with that of SkM* [Figs.~\ref{Ca_def_SkM}(a) and \ref{Ca_def_SkM}(b)]. 
All the states have sizable triaxiality with $|\gamma|$ being comparable with or even larger than that for
the SkM* results.  The configurations 
$[6545]_\nu [5555]_\pi$ and $[6554]_\nu [5555]_\pi$ among 
the four negative-parity excited SD states exhibit positive-$\gamma$
deformation ($\gamma \simeq 10^\circ$)  while the other two configurations
  $[5654]_\nu [5555]_\pi$ and $[5645]_\nu [5555]_\pi$ have negative-$\gamma$ deformation
 which varies from $\gamma \simeq -10^\circ$ at low spins to $\gamma \simeq -60^\circ$
  at the band termination 
  around $I \simeq 17$ -- $19$. This characteristic feature of triaxiality is in good agreement with
  the results of SkM*, and we thus confirm the effect of the orbitals $[200]1/2 (r=+\ii)$  
  and $[200]1/2 (r=-\ii)$ which favors positive- and negative-$\gamma$
  deformations, respectively. We can see also that the same effect accounts for 
  the  sign  and the evolution of the $\gamma$-deformations  
  of the ND configuration   $[6644]_\nu [6644]_\pi$ as well as the ``2p2h'' configurations 
   $[6545]_\nu [6545]_\pi$ and $[5645]_\nu [5645]_\pi$.

Focusing on the shape evolution at higher spins  $I \gtrsim 20$ ($\omega_\mathrm{rot} \gtrsim 2.0$ MeV$/\hbar$), 
one sees that the negative-parity excited SD configurations $[6545]_\nu [5555]_\pi$ and $[6554]_\nu [5555]_\pi$
change their $\beta$-deformation at $I\simeq 23$ and $\simeq 20$
from $\beta\simeq 0.4$ and $\simeq 0.5$, respectively, to  $\beta \simeq 0.6$, 
which is larger than the $\beta$-deformation of the 
reference SD configuration $[5555]_\nu [5555]_\pi$. 
In fact these highly-deformed states correspond to the HD configuration, 
where the hyperintruder $g_{9/2}$ shell is occupied by both a neutron and a proton. 
As we discussed in the case of SkM*, 
the occupation of  $[200]1/2 (r=+\ii)$ orbital in these bands at low spins
induces the shape evolution toward the HD configuration via the level crossing between the 
$[200]1/2 (r=+\ii)$ and $[440]1/2 (r=+\ii)$ orbitals.
At even higher spins $I \gtrsim 27$, we see a difference between SkM* and SkI4, for which  
we did not obtain the expected HD state
with an additional occupation of the $\pi[440]1/2 (r=+\ii)$ orbital. 
Another difference is seen
 in the MD band, which, in the case of SkI4, 
 crosses with the SD bands at spins lower than the crossings between the SD and HD  bands.

Next, we discuss the near-yrast structure of $^{41}$Ca. 
Figure~\ref{Ca_band_SkI}(b) shows the excitation energies for some configurations near the yrast state. 
The ND bands  are the yrast up to terminating spin $I \sim 18$. 
Next lowest at low spins are the positive-parity SD bands
in a signature pair, $[6555]_\nu [5555]_\pi$  and $[5655]_\nu [5555]_\pi$ 
(8p8h+$n$ in Table~\ref{41Ca_config}),
which have configurations with a neutron
in the $[200]1/2 (r=\pm\ii$) orbital added to the reference SD configuration in $^{40}$Ca. 
The negative-parity SD bands such as $[6555]_\nu [6545]_\pi$  and $[6555]_\nu [6554]_\pi$ (7p7h+$n$)
with an additional proton ph-excitation appear above the positive-parity SD bands.
The energy ordering of these bands is essentially the same as that of SkM* [Fig.~\ref{Ca_band_SkM}(b)]
as far as the low spin region $I\lesssim 6 $ is concerned. However, the pair of the positive-parity SD bands
$[6555]_\nu [5555]_\pi$  and $[5655]_\nu [5555]_\pi$ (8p8h+$n$) is well separated from the
negative-parity SD bands, and stay as the lowest-energy SD bands
for a longer spin interval up to $I \simeq 12$ 
as compared with the case of SkM*. 
The trend continues further at higher spins $I \simeq 12$ -- $27$.
These are due to the larger SD gap at $Z=20$ in the SkI4 functional.
This indicates that  the positive-parity  
SD/HD bands $[6555]_\nu [5555]_\pi$  and $[5655]_\nu [5555]_\pi$ might be populated
more strongly in fusion reactions if the reality is close to the results of SkI4 than those of SkM*.

The predicted shape deformation and its evolution 
with $\omega_\mathrm{rot}$, including the band termination and hyperdeformation, shown in 
Figs.~\ref{Ca_def_SkI}(a) and \ref{Ca_def_SkI}(b), are also consistent with the results of SkM* except the following differences:
All the configurations have non-zero and sizable triaxiality with $|\gamma| \gtrsim 10^\circ$ 
at $\omega_\mathrm{rot}=0$, as mentioned above.  Correspondingly $\omega_\mathrm{rot}$-dependence of the triaxiality 
is weak at small rotational frequency although the triaxiality itself is larger than those in SkM*.
This indicates that 
the effect of the occupation in the orbitals $[200]1/2 (r=\pm\ii$) inducing the positive or negative $\gamma$
deformation is stronger than that in SkM*.
We observe also that  the negative-parity bands $[6555]_\nu [5654]_\pi$ and $[5655]_\nu [6554]_\pi$
have definite positive-$\gamma$ deformation ($\simeq +10^\circ$)  although
the different signature members of the orbitals  $[200]1/2 (r=\pm\ii$) are occupied by a neutron and a
proton in these configurations. It suggests  that the orbital $[200]1/2(r = +\ii)$ has the effect stronger than
$[200]1/2(r = -\ii)$.

\section{Summary  \label{summary}}
We have investigated the 
shape evolution of 
the doubly-magic $^{40}$Ca and its neighboring $^{41}$Ca in response to rotation 
in the framework of the nuclear energy-density functional method. 
The cranked Skyrme-Kohn-Sham equation was solved in the 3D lattice to describe 
various types of configurations including the negative-parity excited bands. 
We found that the hyperdeformed (HD) states appear above $I\simeq 25$. 
The occupation of both a neutron and a proton in the rotation-aligned [440]1/2 
orbital brings about the occurrence of the HD states. 
Here, we found that the development of triaxial deformation is important to understand 
the property of the near-yrast bands. 
The configurations in which the $[200]1/2 (r=-\ii)$ orbital is occupied by a neutron and/or a proton 
undergo the evolution of triaxial deformation with negative value with an increase in spin, 
and terminate at the intermediate spins without the band crossing.  
The occupation of the $[200]1/2 (r=+\ii)$ orbital at low spins leads to the realization of the HD states at high spins 
after the band crossings. 
We predicted that this mechanism can be verified in an experimental observation of the positive-parity 
superdeformed (SD) signature-partner bands in $^{41}$Ca 
which has an 8p8h+$n$ configuration with the last neutron occupying the $[200]1/2 (r=+\ii)$ or $[200]1/2 (r=-\ii)$ 
orbital on top of the reference SD configuration $[5555]_\nu [5555]_\pi$ in $^{40}$Ca: 
One of the signature-partner bands with $\alpha=1/2$ $(I=1/2, 5/2, \cdots)$
undergoes band crossings leading to the HD band, 
while the other partner band with $\alpha=-1/2$ $(I=3/2, 7/2, \cdots)$ terminates below $I \simeq 20$. 

\section*{Acknowledgment}
We thank T.~Inakura and T.~Nakatsukasa for valuable discussions, and E.~Ideguchi for communications on the present experimental status. 
This work was supported by the JSPS KAKENHI (Grants No. JP16K17687, No. JP17K05436, and No. JP19K03824) 
and the JSPS-NSFC Bilateral Program for Joint Research Project on 
``Nuclear mass and life for unraveling mysteries of the r-process''. 
The numerical calculations were performed on CRAY XC40 at the Yukawa Institute for Theoretical Physics, Kyoto University, 
and on COMA (PACS-IX) at the Center for Computational Sciences, University of Tsukuba.

\appendix

\section{Calculation of the ground-state energy for an odd-$A$ nucleus}
For an odd-$A$ nucleus with spherical symmetry, 
the solutions of the deformed Skyrme-Kohn-Sham equation are not converged generally 
due to the degeneracy of the single-particle orbitals with different magnetic quantum numbers;  $(2j+1)$-fold degeneracy.  
We encountered the oscillation of the calculated 
$J_z$, expectation value of the $z$-component of total angular momentum operator. 
This is because one can not define the symmetry axis and thus the $J_z$ can not be defined uniquely. 
To resolve the degeneracy of the single-particle orbitals, 
the infinitesimal cranking is introduced:
\begin{equation}
\epsilon^\prime_m = \epsilon - \omega_\mathrm{rot}m, \hspace{1cm} m=\pm \frac{1}{2}\hbar, \pm\frac{3}{2}\hbar, \cdots, \pm j \hbar,
\end{equation}
with $m$ being the magnetic quantum number of the total angular momentum. 
Then, we take the extrapolation to the limit of $\omega_{\rm rot} \to 0$.


%

\bibliographystyle{ptephy}
\bibliography{cranking_ref}

\end{document}